\documentclass[draftclsnofoot,onecolumn,12pt]{IEEEtran}

\usepackage{graphicx,epsf,psfrag}
\usepackage{amsmath,amssymb,amsfonts,verbatim}
\usepackage{mathrsfs}
\usepackage{etoolbox}

\usepackage{latexsym}

\newcommand{\ie}{\emph{i.e.\ }}

\newcommand{\cf}{\emph{cf.\ }}

\newcommand{\beq}{\begin{equation}}
\newcommand{\eeq}{\end{equation}}
\newcommand{\eps}{\epsilon}

\newcommand{\bi}{\begin{itemize}}
\newcommand{\ei}{\end{itemize}}

\newcommand{\calA}{\mathcal{A}}

\newcommand{\calC}{\mathcal{C}}

\newcommand{\calE}{\mathcal{E}}

\newcommand{\calN}{\mathcal{N}}

\newcommand{\calP}{\mathcal{P}}
\newcommand{\calQ}{\mathcal{Q}}
\newcommand{\calR}{\mathcal{R}}
\newcommand{\calS}{\mathcal{S}}

\newcommand{\calV}{\mathcal{V}}
\newcommand{\calW}{\mathcal{W}}
\newcommand{\calX}{\mathcal{X}}
\newcommand{\calY}{\mathcal{Y}}
\newcommand{\calZ}{\mathcal{Z}}

\newcommand{\bx}{\mathbf{x}}
\newcommand{\bX}{\mathbf{X}}
\newcommand{\by}{\mathbf{y}}
\newcommand{\bY}{\mathbf{Y}}

\newcommand{\bbE}{\mathbb{E}}
\newcommand{\bbF}{\mathbb{F}}

\newcommand{\bbP}{\mathbb{P}}

\newcommand{\scR}{\mathscr{R}}

\DeclareMathAlphabet{\mathbsf}{OT1}{cmss}{bx}{n}
\DeclareMathAlphabet{\mathssf}{OT1}{cmss}{m}{sl}

\newcommand{\rvP}{\mathsf{P}}

\newcommand{\rvS}{\mathsf{S}}

\DeclareSymbolFont{bsfletters}{OT1}{cmss}{bx}{n}  
\DeclareSymbolFont{ssfletters}{OT1}{cmss}{m}{n}
\DeclareMathSymbol{\bsfGamma}{0}{bsfletters}{'000}
\DeclareMathSymbol{\ssfGamma}{0}{ssfletters}{'000}
\DeclareMathSymbol{\bsfDelta}{0}{bsfletters}{'001}
\DeclareMathSymbol{\ssfDelta}{0}{ssfletters}{'001}
\DeclareMathSymbol{\bsfTheta}{0}{bsfletters}{'002}
\DeclareMathSymbol{\ssfTheta}{0}{ssfletters}{'002}
\DeclareMathSymbol{\bsfLambda}{0}{bsfletters}{'003}
\DeclareMathSymbol{\ssfLambda}{0}{ssfletters}{'003}
\DeclareMathSymbol{\bsfXi}{0}{bsfletters}{'004}
\DeclareMathSymbol{\ssfXi}{0}{ssfletters}{'004}
\DeclareMathSymbol{\bsfPi}{0}{bsfletters}{'005}
\DeclareMathSymbol{\ssfPi}{0}{ssfletters}{'005}
\DeclareMathSymbol{\bsfSigma}{0}{bsfletters}{'006}
\DeclareMathSymbol{\ssfSigma}{0}{ssfletters}{'006}
\DeclareMathSymbol{\bsfUpsilon}{0}{bsfletters}{'007}
\DeclareMathSymbol{\ssfUpsilon}{0}{ssfletters}{'007}
\DeclareMathSymbol{\bsfPhi}{0}{bsfletters}{'010}
\DeclareMathSymbol{\ssfPhi}{0}{ssfletters}{'010}
\DeclareMathSymbol{\bsfPsi}{0}{bsfletters}{'011}
\DeclareMathSymbol{\ssfPsi}{0}{ssfletters}{'011}
\DeclareMathSymbol{\bsfOmega}{0}{bsfletters}{'012}
\DeclareMathSymbol{\ssfOmega}{0}{ssfletters}{'012}

\newcommand{\hatA}{\hat{A}}

\newcommand{\hath}{\hat{h}}

\newcommand{\tilQ}{\tilde{Q}}

\newcommand{\tilR}{\tilde{R}}

\newcommand{\hats}{\hat{s}}
\newcommand{\hatS}{\hat{S}}

\newcommand{\hatw}{\hat{w}}
\newcommand{\hatW}{\hat{W}}

\newcommand{\hatY}{\hat{Y}}

\newcommand{\tilY}{\tilde{Y}}

\newcommand{\hatZ}{\hat{Z}}

\newcommand{\bars}{\bar{s}}

\newcommand{\barC}{\bar{C}}

\DeclareMathOperator*{\argmax}{arg\,max}
\DeclareMathOperator*{\argmin}{arg\,min}

\newcommand{\qednew}{\nobreak \ifvmode \relax \else
      \ifdim\lastskip<1.5em \hskip-\lastskip
      \hskip1.5em plus0em minus0.5em \fi \nobreak
      \vrule height0.75em width0.5em depth0.25em\fi}

\usepackage{color}
\usepackage{xspace}
\usepackage{algpseudocode}
\usepackage{algorithm}
\usepackage{cite}

\renewcommand{\hatW}{\widehat{W}}
\renewcommand{\hatA}{\widehat{A}}

\newcommand{\ucalN}{\underline{\calN}}
\newcommand{\ucalX}{\underline{\calX}}
\newcommand{\ucalY}{\underline{\calY}}

\newcommand{\ucalC}{\underline{\calC}}
\newcommand{\ucalW}{\underline{\calW}}
\newcommand{\ux}{\underline{x}}
\newcommand{\uy}{\underline{y}}
\newcommand{\uw}{\underline{w}}
\newcommand{\ubx}{\underline{\bx}}
\newcommand{\uby}{\underline{\by}}
\newcommand{\us}{\underline{s}}
\newcommand{\uS}{\underline{S}}
\newcommand{\uY}{\underline{Y}}
\newcommand{\uX}{\underline{X}}
\newcommand{\uW}{\underline{W}}
\newcommand{\ubX}{\underline{\bX}}
\newcommand{\ubY}{\underline{\bY}}
\newcommand{\uhatW}{\underline{\hatW}}
\newcommand{\uhatw}{\underline{\hatw}}
\newcommand{\CC}{\textrm{\textnormal{CC}}}
\newcommand{\AVC}{\textrm{\textnormal{AVC}}}

\allowdisplaybreaks[1]

\newcommand{\xuc}{x^{(\{u\}^c)}}
\newcommand{\Xuc}{X^{(\{u\}^c)}}

\newtheorem{theorem}{Theorem}
\newtheorem{lemma}[theorem]{Lemma}
\newtheorem{proposition}[theorem]{Proposition}
\newtheorem{corollary}[theorem]{Corollary}
\newtheorem{definition}{Definition}

\begin{document}
\title{Equivalence for Networks with Adversarial State}

\author{Oliver Kosut, \IEEEmembership{Member, IEEE} and J{\"o}rg Kliewer, \IEEEmembership{Senior Member, IEEE}
\thanks{O.~Kosut is with the School of Electrical, Computer and Energy Engineering, Arizona State University, Tempe, AZ 85287 USA (e-mail: \hbox{okosut@asu.edu}).}
\thanks{J. Kliewer is with the Department of Electrical and Computer Engineering, New Jersey Institute of Technology, Newark, NJ 07102 USA (email: \hbox{jkliewer@njit.edu}).}
\thanks{This work was presented in part at the 2014 IEEE International Symposium on Information Theory.}
\thanks{This work was supported in part by the
U.S.~National Science Foundation under grants CCF-1439465, CCF-1440014, and CCF-1453718.}
}

\maketitle

\begin{abstract}
We address the problem of finding the capacity of noisy networks with either independent point-to-point compound channels (CC) or arbitrarily varying channels (AVC). These channels model the presence of a Byzantine adversary which controls a subset of links or nodes in the network. We derive equivalence results showing that these point-to-point channels with state can be replaced by noiseless bit-pipes without changing the network capacity region. Exact equivalence results are found for the CC model, and for some instances of the AVC, including all nonsymmetrizable AVCs. These results show that a feedback path between the output and input of a CC can increase the equivalent capacity, and that if common randomness can be established between the terminals of an AVC (either by feedback, a forward path, or via a third-party node), then again the equivalent capacity can increase. This leads to an observation that deleting an edge of arbitrarily small capacity can cause a significant change in network capacity. We also analyze an example involving an AVC for which no fixed-capacity bit-pipe is equivalent.
\end{abstract}

\section{Introduction}
\label{sec:introduction}
One fundamental problem in wireless and wireline networks is to achieve
robustness against active adversaries. A common assumption is to consider
Byzantine adversaries who observe all transmissions, messages, and channel
noise values and interfere with the transmitted signals, i.e., by replacing
a subset of the channel output values or by injecting additional noise to a
specific subset of communication channels or nodes (the adversarial set) in the
network. For example,  for the
adversarial noiseless case both in-network error
correction approaches and capacity results under network coding  have been presented, e.g., in
\cite{Jaggi_etal08,KHEA09,KTT09,KTT10}.

The underlying uncertainty in the network due to the action of the adversary
leads to channels with varying state in the adversarial set
\cite{LN98}. One possible model is to assume that the
corresponding nodes have no knowledge about the  exact channel state, but only that
the state is selected from a finite set. In the
case of a compound channel (CC) \cite{BBT59,Wolf78} the
selected 
state is fixed over the whole transmission of a codeword. In contrast, if
the channel state varies from symbol to symbol in an unknown and arbitrary
manner we have the case of an arbitrarily varying channel (AVC)
           \cite{BBT60,Ahl78,CsiszarNarayan:88IT,CsiszarKorner:Book11}.

\begin{figure}[tb]
\centerline{
\includegraphics[width=.45\textwidth]{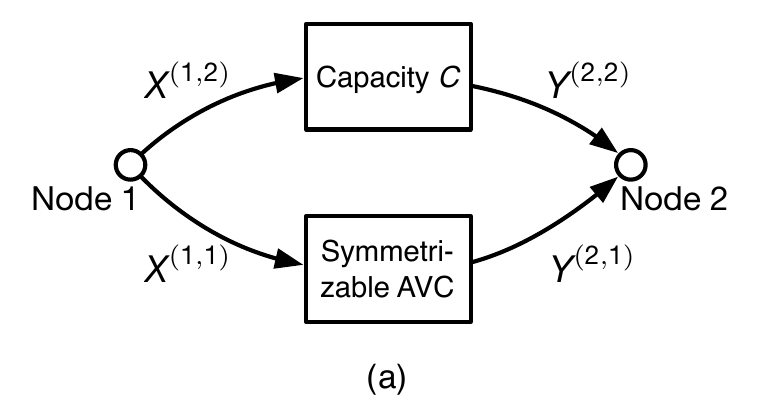}\includegraphics[width=.45\textwidth]{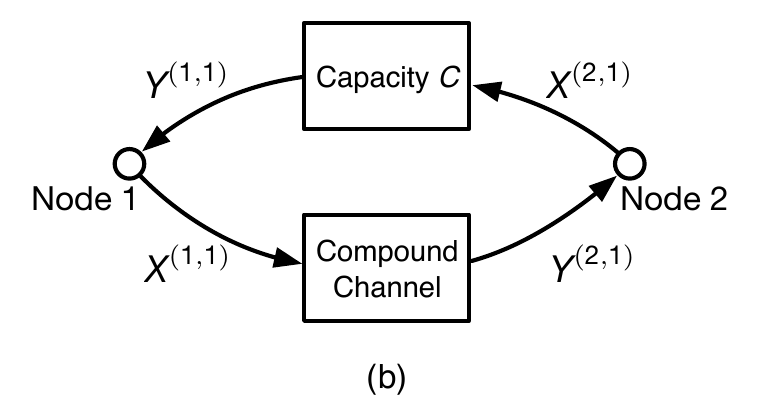}}
\caption{Two-node networks with a capacity $C$ channel and (a) a symmetrizable AVC,
  (b) a CC. In general, the upper channel can be replaced with a single-source
single-sink network having the same rate.}
\vspace{-4ex}
\label{fig:AVC_network}
\end{figure}

Note that the AVC  has a (deterministic) capacity which is either zero or
equals the random coding capacity \cite{Ahl78}. The former case holds
for a symmetrizable AVC, since such a channel can mimic a valid input
sequence in such a way that it is impossible for the decoder to decide on
the correct codeword. Even though transmission is not possible if
such an AVC is considered in isolation, the situation changes in a network
setting, as exemplarily depicted in Fig.~\ref{fig:AVC_network}(a).
In this two-node network, source and destination nodes are connected via two
parallel channels, a (fixed) channel with capacity $C$ and a symmetrizable
AVC. Here, communication over the AVC is possible with a non-zero rate since
common randomness with negligible rate $\epsilon>0$ can be shared between
both nodes \cite{Ahl78,CsiszarNarayan:88IT,CsiszarKorner:Book11} via the
upper channel in Fig.~\ref{fig:AVC_network}(a). In a more general setup,
in  Fig.~\ref{fig:AVC_network} this channel can be replaced with a single-source
single-sink network of positive rate $C$.

In the following we consider the problem of reliable communication over a network of
independent noisy point-to-point channels in the presence of active
adversaries. A subset of the channels either consists of AVCs or CCs. This is in
contrast to the model in \cite{BEH11}, where the action of the adversary is
directly modeled by injecting an arbitrary vector to the network edges in
the adversarial set. By building on the results in \cite{Equivalence} we
identify cases where 
the adversarial capacity of the network equals the capacity of
another network in which each channel is replaced by a noise-free
bit-pipe. For a CC, the bit-pipe has capacity equal to the standard CC
capacity if there is no feedback path from the output to the input; if there
is, then the equivalent bit-pipe has higher capacity, because the state can
be estimated at the output and relayed back to the input (see
Fig.~\ref{fig:AVC_network}(b)). For an AVC, the equivalent bit-pipe has
capacity equal to the random coding capacity if it is possible to establish
common randomness between the input and output. This can be accomplished if
any of the following hold: (i) the AVC is non-symmetrizable, (ii) there is a
parallel forward path as in Fig.~\ref{fig:AVC_network}(a), (iii) there is a
feedback path as for the CC in Fig.~\ref{fig:AVC_network}(b), or (iv) a
third-party node can transmit to both the input and output nodes. If none of
these hold, it appears to be difficult to obtain an equivalence result, as
the strong converse does not hold for symmetrizable AVCs. Indeed, we
illustrate in Sec.~\ref{sec:example} that there exist AVC networks in which
no equivalent bit-pipe with fixed capacity exists.

{These observations are related to the concept of super-activation
  \cite{Duan09} which for two channels $\mathcal{C}_1$ and $\mathcal{C}_2$ is
  defined by the observation  that these channels can only be used for
  reliable communication if they are used jointly, but not in
  isolation. Super-activation has for example been studied in the context of
  arbitrarily-varying wiretap
  channels \cite{BSP15,NWB16}, where it has been
  shown that there exist pairs of symmetrizable  arbitrarily-varying wiretap
  channels which can be super-activated.}

The structure of the paper is as follows. In Sec.~\ref{sec:model}, we formally introduce the problem for both CC and AVC models. In Sec.~\ref{sec:stacked} we describe the concept of stacked networks, introduced in \cite{Equivalence}, and state two preliminary lemmas. In Sec.~\ref{sec:training}, we introduce a lemma demonstrating that training sequences can be used for the CC model to reliably estimate the channel state. {In Sec.~\ref{sec:avc_randomness}, we prove a lemma for the AVC model showing that having access to unlimited shared randomness among certain sets of nodes does not change the capacity region.} In Sec.~\ref{sec:positive}, given a channel model and a pair of nodes $u$ and $v$, we determine whether it is possible to transmit information at any positive rate from $u$ to $v$. These results will be used in the equivalence results for both state models: for the CC model, to determine whether feedback is possible, and for the AVC model, whether common randomness can be established (\cf Fig.~\ref{fig:AVC_network}). In Sec.~\ref{sec:cc} we present our main equivalence results for the CC model, and in Sec.~\ref{sec:avc} for the AVC model. In Sec.~\ref{sec:example} we analyze an example AVC network that we show has no equivalent bit-pipe. In Sec.~\ref{sec:edge_removal} we relate our results to the edge removal problem, which has proved difficult for state-less networks but we prove has a simple solution for both CC and AVC models. We conclude in Sec.~\ref{sec:conclusion}.

\section{Model}\label{sec:model}

Consider a network of nodes $\calV:=\{1,\ldots,m\}$ with state, given by
\beq\label{eq:model}
\calN=\left(\prod_{v=1}^m \calX^{(v)},\calS,p(\by|\bx,s),\prod_{v=1}^m\calY^{(v)}\right).
\eeq
Herein, $\calX^{(v)}$ and $\calY^{(v)}$ denote the input and output alphabets
of the  node $v$ and $\calS$ the set of network states, respectively. 
This network may represent either a CC or an AVC model. These both assume
that the state is chosen not randomly but adversarially; in the CC model the
adversary chooses a single state $s\in\calS$ that remains constant
throughout the code block, whereas in the AVC model the adversary chooses an
arbitrary state sequence $s^n\in\calS^n$. 
{We assume that the adversary is blind,}
  i.e., that it does not know the transmitted messages,
  but only the employed codebooks.
     In this paper we are interested in both
CC and AVC problems, but only one at a time. Studying networks with both CC-type state and AVC-type state is beyond our scope. 

{We also assume that nodes may use private randomness, independently generated at each node, in their coding operations. This will be important for our results on the AVC achievability arguments, in which nodes generated random quantities and transmit them across the network. In our model we allow each node an unlimited amount of private randomness (in particular, a uniform random variable on the interval $[0,1]$), although our achievability arguments require no more than $O(\log n)$ bits of private randomness are required at each node. Note that private randomness is quite different from shared randomness, which is a significant asset that trivializes many AVC problems; we do \emph{not} assume that any shared randomness is available in this model. In Sec.~\ref{sec:avc_randomness} we show that certain forms of shared randomness have no effect on the capacity region of the AVC model, but this is not true for unrestricted shared randomness.}

{
We further assume that there is an independent point-to-point channel from node 1 to node 2 with independent state. That is, $\calX^{(1)}=\calX^{(1,0)}\times\calX^{(1,1)}$, $\calY^{(2)}=\calY^{(2,0)}\times\calY^{(2,1)}$, $\calS=\calS^{(0)}\times\calS^{(1)}$, and
\beq
p(\by|\bx,s)=p(\by^{(0)}|\bx^{(0)},s^{(0)}) p(y^{(2,1)}|x^{(1,1)},s^{(1)})
\eeq
where  $x^{(1,1)}\in\calX^{(1,1)}$, $y^{(2,1)}\in\calY^{(2,1)}$, and
$s^{(1)}\in\calS^{(1)}$ represent the input, output, and state respectively
for the point-to-point channel, and
$\bx^{(0)}\in\calX^{(1,0)}\times\prod_{v\ne 1} \calX^{(v)}$,
$\by^{(0)}\in\calY^{(2,0)}\times\prod_{v\ne 2}\calY^{(v)}$, and
$s^{(0)}\in\calS^{(0)}$ represent the input, output, and state respectively
for the remainder of the network. {This decomposition is visualized in
  Fig.~\ref{fig:network}.}
\begin{figure}[tb]
\centerline{\includegraphics[scale=0.6]{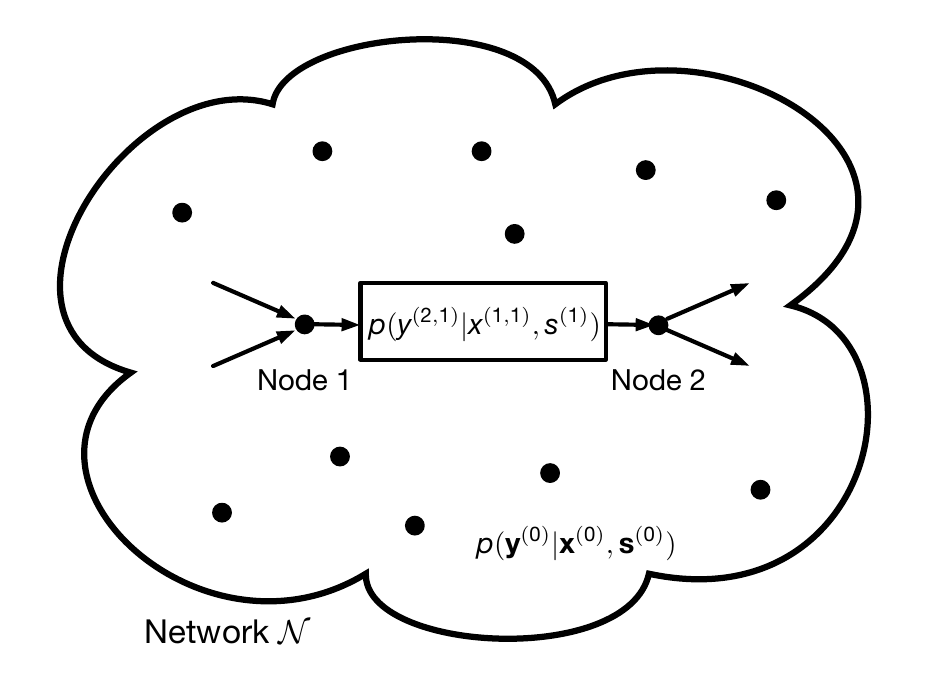}}
\caption{{Decomposition of a network $\mathcal{N}$ into a point-to-point
  channel between node 1 and 2 with conditional pmf
  $p(y^{(2,1)}|x^{(1,1)},s^{(1)})$ and the channels specified by the rest of the network with pmf $p(\by^{(0)}|\bx^{(0)},s^{(0)})$.}}
\label{fig:network}
\end{figure}
The point-to-point channel itself is given by
\beq
\calC=(\calX^{(1,1)},\calS^{(1)},p(y^{(2,1)}|x^{(1,1)},s^{(1)}),\calY^{(2,1)}).
\eeq
}

{
Our main goal is to relate the capacity region of $\calN$ to that when point-to-point channel $\calC$ is replaced by a noiseless link of fixed capacity (i.e. a bit-pipe). In particular, for any $R\ge 0$, let $\calN^R$  be the network in which $\calC$ is replaced by a rate-$R$ noiseless (and
state-less) bit-pipe $\calC^R$ given by
{
\[
\calC^R=(\{0,1\}^R,\delta(y^{(2,1)}-x^{(1,1)}),\{0,1\}^R).
\]
With other words, the noiseless bit-pipe
of capacity $R$ transmits  $\lfloor nR\rfloor$ bits over each block of $n$
channel uses with zero error probability for any integer $n\ge 1$. }
}

In general, CCs and AVCs can be quite pathological, so we assume that
alphabets $\calX^{(v)}$, $\calS$, and $\calY^{(v)}$ are all finite
sets. Most of our results apply for more general alphabets under mild
regularity conditions, but to avoid edge cases and complications we restrict
ourselves to finite alphabets. We believe that the interesting consequences of the CC and AVC network models are captured with finite alphabets models, and that the complications that arise for general alphabets are unlikely to make a difference in practice.

\emph{Notation}: Let $[k]=\{1,\ldots,k\}$. A rate vector $\calR$ consists of multicast rates $R^{(\{v\}\to U)}$ from each source node $v$ to each destination set $U\subseteq\calV$. With a singleton destination set $U=\{u\}$, we sometimes write simply $R^{(v\to u)}$. For each $(v,U)$ pair, there is a message $W^{(\{v\}\to U)}\in\calW^{(\{v\}\to U)}=[2^{nR^{(\{v\}\to U)}}]$. Let $W^{(V\to *)}$ denote the vector of all messages originating at nodes $v\in V$, and let $\calW^{(V\to *)}$ denote the corresponding message set. Also let $W$ denote the vector of all messages. For a set $\calA\subset\calV$, we write $X^{(\calA)}=(X^{(v)}:v\in\calA)$, and similarly for $Y^{(\calA)}$. We also write $\bx$ for $X^{(\calV)}$ and $\by$ for $Y^{(\calV)}$, as in \eqref{eq:model}. {For each node $v$, the private randomness generate at node $v$ is given by a random variable $Q_v$ drawn uniformly from the interval $[0,1]$.}

A blocklength-$n$ solution $\rvS(\calN)$ for network $\calN$ is given by:
{
\begin{itemize}
\item for each $v\in\calV$ and $t\in[n]$, an encoding function
\beq\label{eq:encoding_function}
X_t^{(v)}:(\calY^{(v)})^{t-1}\times \calW^{(\{v\}\to *)}\times [0,1]\to \calX^{(v)}
\eeq
by which node $v$ determines channel input symbol $X_t^{(v)}$ given previously received data $Y^{(v)}_{1:t-1}$, messages $W^{(\{v\}\to *)}$, and private randomness $Q_v$
\item for each $(v,U)$ pair and each $u \in U$, a decoding function
\beq\label{eq:decoding_function}
\hatW^{(\{v\}\to U),u}:(\calY^{(u)})^n\times \calW^{(\{u\}\to *)}\times[0,1]\to \calW^{(\{v\}\to U)}\cup\{e\}
\eeq
by which node $u$ determines a message estimate $\hatW^{(\{v\}\to U),u}$ of $W^{(\{v\}\to U)}$ given received data $Y^{(u)}_{1:n}$, messages $W^{(\{u\}\to *)}$, and private randomness $Q_u$. Here, $e$ is a special symbol that denotes declaring an error.
\end{itemize}
}
Let $\hatW$ be the complete vector of message estimates, and denote by  $\{\hatW\ne W\}$ the event that at least one message is incorrectly decoded. Note that the probability of this event depends on the state sequence $S^n$.

\begin{definition}
The CC-capacity region $\scR_{\CC}(\calN)$ of network $\calN$ is given by the closure of the set of rate vectors $\calR$ for which there exists a sequence of blocklength-$n$ solutions for which
\beq
\max_{s\in\calS} \Pr(\hatW\ne W|S^n=(s,s,\ldots,s))\to 0.
\eeq
\end{definition}

\begin{definition}
The AVC-capacity region $\scR_{\AVC}(\calN)$ of network $\calN$ is given by the closure of the set of rate vectors $\calR$ for which there exists a sequence of blocklength-$n$ solutions for which
\beq
\max_{s^n\in\calS^n} \Pr(\hatW\ne W|S^n=s^n)\to 0.
\eeq
\end{definition}

It is easy to see that neither $\scR_{\CC}(\calN)$ nor $\scR_{\AVC}(\calN)$ change if the state is allowed to be randomized instead of deterministic, as long as this random choice is independent of the message and the operation of the channel, and for the CC model the state is fixed across the coding block.

Our goal is to prove achievability-type results of the form $\scR(\calN^R)\subseteq \scR(\calN)$ and converse-type results of the form $\scR(\calN)\subseteq \scR(\calN^R)$ for both CC and AVC models.

\section{Stacked Networks}\label{sec:stacked}

We adopt the notion from \cite{Equivalence} of \emph{stacked networks},
wherein we denote by $\underline{\calN}$ a network with $N$ independent
copies of the network $\calN$. Each copy (layer) contains an instance of every channel input and every channel output, all operating independently\footnote{With the exception that in the CC model, the state is constant across all layers of the network and all time.}. Underlines denote stacked variables and vectors, and the argument $\ell$ refers to layer $\ell$, where $\ell\in[N]$. That is, $\underline{X}^{(v)}(\ell)$ is the symbol transmitted by node $v$ in layer $\ell$, and $\uY^{(v)}(\ell)$ is the symbol received by node $v$ in layer $\ell$. Moreover, we denote $\uX^{(v)}=(\uX^{(v)}(\ell):\ell\in[N])$ and similarly for $\uY^{(v)}$. The corresponding alphabets are given by $\ucalX^{(v)}$, etc. Message sets are correspondingly increased by a factor of $N$; that is, $\ucalW^{(\{v\}\to U)}=(\calW^{(\{v\}\to U)})^N$. Rates are therefore defined by $R^{(\{v\}\to U)}=|\ucalW^{(\{v\}\to U)}|/(nN)$.

We need to differentiate between the CC and AVC models for stacked networks, because for the CC model the state remains constant across time and across layers, whereas for the AVC model the state may vary between layers. For the CC model, the distribution of channel outputs $\underline{\bY}=(\uY^{(v)}:v\in\calV)$ given channel inputs $\underline{\bX}=(\uX^{(v)}:v\in\calV)$ and state $s\in\calS$ is
\beq
p(\underline{\by}|\underline{\bx},s)=\prod_{\ell=1}^N p(\uby(\ell)|\ubx(\ell),s)
\eeq
where $\ubX(\ell)$ and $\ubY(\ell)$ are  the vectors of transmitted and received symbols respectively in layer $\ell$. For the AVC model, there is a different state in each layer denote $\uS(\ell)$ for layer $\ell$. The distribution of $\ubY$ given $\ubX$ and state vector $\uS=(\uS(\ell):\ell\in[N])$ is
\beq
p(\uby|\ubx,\us)=\prod_{\ell=1}^N p(\uby(\ell)|\ubx(\ell),\us(\ell)).
\eeq

Solutions for stacked networks are defined similarly to those for unstacked
networks, the only difference being that each coding function has access to
all stacks from prior time instances. In particular, the transmitted symbols
for all layers at node $v$ and time $t$ are determined by the causal encoding function 
\beq
\uX_t^{(v)}:(\ucalY^{(v)})^{t-1}\times \ucalW^{(\{v\}\to *)}\to \ucalX^{(v)}
\eeq
and the decoding function for message $\uW^{(\{v\}\to U)}$ at node $u\in U$ is given by
\beq
\uhatW^{(\{v\}\to U),u}:(\ucalY^{(u)})^n\times \ucalW^{(\{u\}\to *)}\to \ucalW^{(\{u\}\to U)}\cup\{e\}.
\eeq
Note that node $v$ has access to its received symbols and messages in all layers when deciding its transmissions. The capacity regions for the stacked networks $\scR_{\CC}(\ucalN)$ and $\scR_{\AVC}(\ucalN)$ are defined analogously as above for unstacked networks.

The following two preliminary lemmas are simple extensions of Lemmas~1 and~4 respectively from \cite{Equivalence} to include state.

\begin{lemma}\label{lemma:stacked}
For any network $\calN$, $\scR_{\CC}(\calN)=\scR_{\CC}(\underline\calN)$ and $\scR_{\AVC}(\calN)=\scR_{\AVC}(\underline\calN)$.
\end{lemma}
\begin{IEEEproof}
{The proof for the two state models are largely the same, so we describe
  them both simultaneously and discuss differences  only when they arise. We
  first prove $\scR_{\CC}(\calN)\subseteq\scR_{\CC}(\underline\calN)$ and
  $\scR_{\AVC}(\calN)\subseteq\scR_{\AVC}(\underline\calN)$. Consider any
  rate $\calR$ in the interior of the capacity region for $\calN$, and we
  prove that $\calR$ is achievable for $\underline\calN$. This is sufficient
  because of the closure operation in the definition of the capacity
  regions. Given any $\lambda>0$, for $n$ sufficiently large there exists a
  blocklength-$n$ solution $\rvS(\calN)$ on network $\calN$ with rate
  $\calR$ and probability of error $\lambda/N$. We construct a solution for
  stacked network $\underline\calN$ by repeating $\rvS(\calN)$ identically
  and independently on each layer of $\underline\calN$. Note that the
  independence refers to the encoding operations at the nodes, wherein
  layers are independent of each other, but not necessarily to the input and
  output random variables at different layers, which may be made dependent via the adversarial state. However, it is still the case that the probability of error for each layer is at most $\lambda/N$, because each layer looks like an ordinary CC or AVC-type model.\footnote{{In the CC model, the adversary is restricted to maintain a constant state across layers, but this assumption is not necessary for this direction of proof.}} Thus, by the union bound, the probability of error for the stacked solution is at most $\lambda$.}

{We now prove $\scR_{\CC}(\calN)\supseteq\scR_{\CC}(\underline\calN)$ and
  $\scR_{\AVC}(\calN)\supseteq\scR_{\AVC}(\underline\calN)$. Given any
  blocklength-$n$ solution on $\underline\calN$, it may be ``unraveled'' to
  form a blocklength-$nN$ solution on $\calN$ with identical rate and
  probability of error. In particular, the symbols transmitted at time $t$
  by the $N$ layers of $\underline\calN$ are transmitted at times
  $(t-1)N+1,\ldots,tN$ on $\calN$. Thus causality is maintained at each
  node. For the AVC model, the same unraveling operation forms an
  equivalence between state sequences selections for the length-$n$ solution
  on $\underline\calN$ and the length-$nN$ solution on $\calN$. Thus the
  worst case probability of error is unchanged. For the CC model, since the
  state is required to be constant across layers in $\underline\calN$, {the
  state selection is unchanged and fixed over the blocklength-$nN$
  solution. This means that} again the 
  state selections are equivalent between the two models, so the probability of error is unchanged.
}
\end{IEEEproof}

\begin{lemma}\label{lemma:continuity}
The capacity regions $\scR_{\CC}(\calN^R)$ and $\scR_{\AVC}(\calN^R)$ are continuous in $R$ for all $R>0$.
\end{lemma}
\begin{IEEEproof}
We employ a very similar proof technique as that of Lemma~4 in \cite{Equivalence}. By Lemma~\ref{lemma:stacked}, it is equivalent to prove continuity for $\scR_{\CC}(\underline\calN^R)$ and $\scR_{\AVC}(\underline\calN^R)$. Fix any $\delta\in (0,R)$ and rate vector $\calR\in\text{int}(\scR_{\CC}(\underline\calN^{R+\delta}))$ (resp. $\calR\in\text{int}(\scR_{\AVC}(\underline\calN^{R+\delta}))$). Assume that $\ucalN^{R+\delta}$ has $N$ layers. Let $\ucalN^{R-\delta}$ be an $N'$-fold stacked network with
\beq\label{eq:N_prime}
N'(R-\delta)\ge N(R+\delta).
\eeq
For all $\lambda>0$, there exists solution $\rvS(\ucalN^{R+\delta})$ with rate vector $\calR$ and probability of error $\lambda$. We define a solution $\rvS(\ucalN^{R-\delta})$ based on $\rvS(\ucalN^{R+\delta})$ as follows. Use precisely the same coding operations aside from the bit-pipe $\ucalC^{R+\delta}$ for the first $N$ layers of the stack, and send the $\lfloor N(R+\delta)\rfloor$ bits to be sent across $\ucalC^{R+\delta}$ instead across the bit-pipe $\ucalC^{R-\delta}$. This can be done because of \eqref{eq:N_prime}. Note that the resulting rate vector for $\rvS(\ucalN^{R-\delta})$ is
\beq
\calR'=\frac{\calR N}{N'}>\calR\frac{N}{N(R+\delta)/(R-\delta)+1}.
\eeq
Thus the difference between $\calR$ and $\calR'$ vanishes as $N\to\infty$ and $\delta\to 0$.

Recall that for the CC-model (resp. AVC-model), the state does not affect operation of the bit-pipes. Meanwhile, as the rest of the network is operated identically in the two solutions---aside from the $N'-N$ unused layers in the solution on $\ucalN^{R-\delta}$---the effect of the state is precisely the same. Thus the modified solution on $\ucalN^{R-\delta}$ has precisely the same probability of error $\lambda$. Therefore $\calR'\in\scR_{\CC}(\ucalN^{R-\delta})$ (resp. $\calR'\in\scR_{\AVC}(\ucalN^{R-\delta})$).
\end{IEEEproof}

\section{Compound Channel Training Lemma}\label{sec:training}

The following lemma will be used several times in CC results. It asserts that CC states can be estimated using training sequences.

\begin{lemma}\label{lemma:training}
Fix a point-to-point CC $(\calX,\calS,p(y|x,s),\calY)$. {For any input sequence $x_{1:n}\in\calX^n$ and output sequence $y_{1:n}\in\calY^n$, define the set of maximum likelihood state estimates as
\begin{equation}
\hat\calS(x_{1:n},y_{1:n})=\{\hats\in\calS: p(y_{1:n}|x_{1:n},\hats)=\max_{s'\in\calS}\, p(y_{1:n}|x_{1:n},s')\}.
\label{eq:Shat}
\end{equation}
For any state $s\in\calS$, let the set of states equivalent to $s$ be\footnote{{In many cases, each state induces a distinct channel distribution, so we would have $\bar\calS(s)=\{s\}$. However, there are important scenarios when this is not the case, such as when $S$ is the full network channel state, and the channel from $X$ to $Y$ represents just part of the overall network channel model. Different states might induce the same behavior from $X$ to $Y$ but different behaviors elsewhere in the network. For example, consider two BSCs and a ternary network state $S$ such that the crossover probabilities of the two channels are $(0,0)$ if $S=0$, $(0,1)$ if $S=1$, or $(1,0)$ if $S=2$. Thus $S=0$ and $S=1$ induce exactly the same behavior in the first channel, but are materially different when considering the entire network.}}
\begin{equation}
\bar\calS(s)=\{\bars\in\calS: p(y|x,\bars)=p(y|x,s)\text{ for all
           }x\in\calX,y\in\calY\}.
\label{eq:Sbar}
\end{equation}
Then, for any $s\in\calS$,
\beq\label{eq:training_conclusion}
\lim_{n\to\infty}\Pr\big(\hat\calS(\alpha_{1:n},Y_{1:n})\ne \bar\calS(s)\big)=0
\eeq 
where $\alpha_{1:n}$ is a random training sequence drawn uniformly i.i.d. from $\calX^n$, and $Y_{1:n}\sim p(y_{1:n}|\alpha_{1:n},s)$.}
\end{lemma}

\begin{IEEEproof}
{Fix $s\in\calS$.} Note that if {$\bars\in\bar\calS(s)$}, then by definition all probabilities for {$\bars$} are identical to those for $s$, so   {$\bars\in\hat\calS(x_{1:n},y_{1:n})$} if and only if {$s\in\hat\calS(x_{1:n},y_{1:n})$}. Thus, to prove \eqref{eq:training_conclusion} we need to show that with probability approaching 1, 
\beq\label{eq:s_proof_goal}
p(Y_{1:n}|\alpha_{1:n},s)>p(Y_{1:n}|\alpha_{1:n},s')\text{ for all }s'\in{\bar\calS(s)^c}
\eeq
where {$\bar\calS(s)^c=\calS\setminus\bar\calS(s)$}.

{Note that $\hat\calS(\alpha_{1:n},Y_{1:n})$ consists of the set of $\hats$ that minimize
\beq
-\frac{1}{n}\sum_{t=1}^n \log p(Y_t|\alpha_t,\hats).
\eeq}
For any $s'\in\calS$, the quantities $-\log p(Y_t|\alpha_t,s')$ are i.i.d.~with expected value
\beq\label{eq:expected_value}
\frac{1}{|\mathcal{X}|}
\sum_{x,y} - p(y|x,s)\log p(y|x,s')
=H(Y|X,S=s)+\Delta_{s,s'}
\eeq
where
\beq
\Delta_{s,s'}:=\sum_{x\in\calX}\frac{1}{|\calX|} D\big(p(y|x,s)\| p(y|x,s')\big).
\eeq
Note that $\Delta_{s,s'}=0$ if and only if $s'\in\bar\calS{(s)}$. Let $\delta_s=\min\{\Delta_{s,s'}:\Delta_{s,s'}>0\}$. We have $\delta_s>0$ since $\calS$ is finite. Let $\tau_s=H(Y|X,S=s)+\delta_s/2$.
{
Hence
\begin{align}
\bbE\left[-\log p(Y_t|\alpha_t,s)\right]&>\tau_s\\
\bbE\left[-\log p(Y_t|\alpha_t,s')\right]&<\tau_s\text{ for any }s'\in\bar\calS{(s)}^c.
\end{align}
Thus, by the Law of Large Numbers,
\begin{align}
\Pr\left( -\frac{1}{n}\sum_{t=1}^n \log p(Y_t|\alpha_t,s) >\tau_s\right)&\to 1\label{eq:s_good}\\
\Pr\left(-\frac{1}{n}\sum_{t=1}^n \log p(Y_t|\alpha_t,s') < \tau_s\text{ for all }s'\in\bar\calS{(s)}^c\right)&\to 1.\label{eq:s_bad}
\end{align}
Therefore \eqref{eq:s_proof_goal} holds with probability approaching 1.
}
\end{IEEEproof}

{
\section{Arbitrarily Varying Shared Randomness Lemma}\label{sec:avc_randomness}
}

{
A key element of proving AVC equivalence results, and indeed of many existing AVC results, is the role of shared randomness between nodes. This is the essence of the difference between the classical deterministic and random coding models for the point-to-point AVC, and so one may ask exactly when does having shared randomness between nodes change or not change the capacity region. In this section, we prove a generic lemma stating that the capacity region of an AVC network does not change if {certain groups of nodes} have access to shared randomness. This lemma will be used several times in proving our equivalence results. For the no adversary model, it was shown in \cite{Xiang2014} that the capacity region for average probability of error does not change even if all nodes have access to a single infinite entropy source of common randomness (modeled as a uniform random variable on the unit interval). This strong result does not hold for the AVC model, but, as stated below, for a given node $v$, if node $v$ is allowed to share an infinite entropy source of randomness with all other nodes to which it can communicate at any positive rate, then the capacity region does not change.
}

{
The proof is a generalization of the random code reduction Lemma~12.8 from \cite{CsiszarKorner:Book11}. This lemma proves that for the point-to-point AVC an arbitrary amount of shared randomness between encoder and decoder can be reduced to an asymptotically negligible amount (in particular, $O(\log n)$ bits). This leads to Theorem~12.11 of \cite{CsiszarKorner:Book11}, stating that the capacity of an AVC is either 0 or the random coding capacity, because if the capacity is positive, than a small amount of shared randomness can be set up, and thus full shared randomness can be simulated. We use essentially the same technique here.
}

{
To be precise, we define the following variant on our coding model.
\begin{definition}\label{def:shared_randomness}
Let $\tilde\scR_{\text{AVC}}(\calN)$ be the capacity region for the AVC network $\calN$ for the following shared randomness coding model. For each node $v$, let ${\tilQ}_v$ be a uniform random variable on the interval $[0,1]$, independent from each other, from the messages, from channel noise, and from the state sequence. Assume ${\tilQ}_v$ is available at node $v$ and at all nodes $u$ for which there exists a rate vector $\calR\in\scR_{\text{AVC}}(\calN)$ with $R^{(v\to u)}>0$. 
\end{definition}
\begin{lemma}\label{lemma:avc_randomness}
For any network $\calN$, $\tilde\scR_{\text{AVC}}(\calN)=\scR_{\text{AVC}}(\calN)$.
\end{lemma}
}

{
Before proving Lemma~\ref{lemma:avc_randomness}, we need the following lemma, which is the essence of the random code reduction.
\begin{lemma}\label{lemma:derandom}
Let $Q$ and $Z$ be independent random variables with (not necessarily finite) alphabets  $\calQ$ and $\calZ$ respectively. Let $f(q,z,s^n)\in[0,1]$ be a function defined for $q\in\calQ$, $z\in\calZ$ and $s^n\in\calS^n$. Suppose for some $\eta>0$,
\beq\label{eq:eta_assumption}
\bbE f(Q,Z,s^n)\le \eta\text{ for all }s^n\in\calS^n.
\eeq
Then for sufficiently large $n$, there exist $q_1,\ldots,q_{n^2}\in\calQ$ such that
\beq\label{eq:derandom}
\frac{1}{n^2} \sum_{j=1}^{n^2} \bbE f(q_j,Z,s^n)\le 2\eta\text{ for all }s^n.
\eeq
\end{lemma}
}
\begin{IEEEproof}
{
Let $Q_1,\ldots,Q_{n^2}$ be i.i.d. random variables with the same distribution as $Q$, all independent of $Z$. We have
\begin{align}
&\bbP\left(\frac{1}{n^2} \sum_{j=1}^{n^2} \bbE[f(Q_j,Z,s^n)|Q_j]>2\eta\text{ for any }s^n\right)\label{eq:derandom1}
\\&\le \sum_{s^n} \bbP\left(\frac{1}{n^2} \sum_{j=1}^{n^2} \bbE[f(Q_j,Z,s^n)|Q_j]>2\eta\right)\label{eq:derandom2}
\\&= \sum_{s^n} \bbP\left(2^{\sum_{j=1}^{n^2}  \bbE[f(Q_j,Z,s^n)|Q_j]}> 2^{n^2 2\eta}\right)\label{eq:derandom3}
\\&\le  \sum_{s^n} 2^{-n^2 2\eta}\, \bbE 2^{\sum_{j=1}^{n^2}  \bbE[f(Q_j,Z,s^n)|Q_j]}\label{eq:derandom4}
\\&=\sum_{s^n} 2^{-n^2 2\eta} \left( \bbE 2^{\bbE[f(Q,Z,s^n|Q)]}\right)^{n^2}\label{eq:derandom5}
\\&\le \sum_{s^n} 2^{-n^2 2\eta} (1+\bbE f(Q,Z,s^n))^{n^2}\label{eq:derandom6}
\\&\le |\calS|^n  2^{-n^2 2\eta} (1+\eta)^{n^2}\label{eq:derandom7}
\\&\le |\calS|^n 2^{-n^2 \eta (2-\log e)}\label{eq:derandom8}
\end{align}
where \eqref{eq:derandom2} follows from the union bound, \eqref{eq:derandom4} from Markov's inequality, \eqref{eq:derandom5} from the fact that $Q_{j}$ for $j\in[n^2]$ are i.i.d. with the same distribution as $Q$, \eqref{eq:derandom6} follows from the fact that $f(q,z,s^n)\in[0,1]$ and $2^x\le 1+x$ for any $x\in[0,1]$, \eqref{eq:derandom7} follows from the assumption in \eqref{eq:eta_assumption}, and \eqref{eq:derandom8} follows because $1+\eta\le e^\eta$. As $2>\log e$, the quantity in \eqref{eq:derandom8} is vanishing in $n$. Thus, for sufficiently large $n$ the probability in \eqref{eq:derandom1} is strictly less than 1, meaning there exists at least one set of constants $\{q_{j}\}_{j\in[n^2]}$ satisfying \eqref{eq:derandom}.
}
\end{IEEEproof}

\begin{IEEEproof}[Proof of Lemma~\ref{lemma:avc_randomness}]
It is obvious that $\scR_{\text{AVC}}(\calN)\subseteq \tilde\scR_{\text{AVC}}(\calN)$. To prove $\tilde\scR_{\text{AVC}}(\calN)\subseteq\scR_{\text{AVC}}(\calN)$, let $\calR$ be a rate vector in the interior of $\tilde\scR_{\text{AVC}}(\calN)$, and we prove that $\calR\in\scR_{\text{AVC}}(\calN)$. 
For sufficiently large $n$ there exists an $n$-length solution $\rvS(\calN)$ for the random coding model with rate $\calR$ and probability of error $2^{-m}\eps$. Given $q_1,\ldots,q_m\in[0,1]$ and $s^n\in\calS^n$, let $e(q_1,\ldots,q_m,s^n)$ be the probability of error for $\rvS(\calN)$ conditioned on ${\tilQ}_i=q_i$ for $i\in\calV$, and $S^n=s^n$. Note that this quantity is averaged over the random choice of messages, the random channel noise, {and any private randomness}. Thus
\beq\label{eq:random_error}
\bbE e({\tilQ}_1,\ldots,{\tilQ}_m,s^n)\le 2^{-m} \eps\text{ for all }s^n.
\eeq
We next prove that there exist $q_{ij}\in[0,1]$ for $i\in\calV$ and $j\in[n^2]$ such that
\beq\label{eq:total_derandom}
\frac{1}{n^{2m}} \sum_{j_1,\ldots,j_m\in[n^2]} e(q_{1j_1},\ldots, q_{mj_m},s^n)
\le \eps\text{ for all }s^n.
\eeq

We now apply Lemma~\ref{lemma:derandom} $m$ times to the initial random coding probability of error in \eqref{eq:random_error}. In particular, by \eqref{eq:random_error}, applying Lemma~\ref{lemma:derandom} with particularizations $e\to f$, ${\tilQ}_1\to Q$, and $({\tilQ}_2,\ldots,{\tilQ}_m)\to Z$, there exists $q_{1j}\in[0,1]$ for $j\in[n^2]$ where
\beq\label{eq:Q1}
\frac{1}{n^2} \sum_{j=1}^{n^2} \bbE e(q_{1j},{\tilQ}_2,\ldots,{\tilQ}_m)\le 2^{-m+1}\eps\text{ for all }s^n.
\eeq
Let $A_1$ be a random variable uniformly distributed on $\{q_{11},\ldots,q_{1n^2}\}$. Thus \eqref{eq:Q1} may be rewritten
\beq
\bbE e(A_1,{\tilQ}_2,\ldots,{\tilQ}_m)\le 2^{-m+1}\eps\text{ for all }s^n.
\eeq
Now applying Lemma~\ref{lemma:derandom} again with particularizations $e\to f$, ${\tilQ}_2\to Q$, and $(A_1,{\tilQ}_3,\ldots,{\tilQ}_m)\to Z$ allows us to conclude that there exist $q_{2j}\in[0,1]$ for $j\in[n^2]$ such that
\beq
\frac{1}{n^2} \sum_{j=1}^{n^2} \bbE e(A_1,q_{2j},{\tilQ}_3,\ldots,{\tilQ}_m)\le 2^{-m+2}\eps\text{ for all }s^n.
\eeq
Repeating this argument $m$ times proves \eqref{eq:total_derandom}.

We now construct a  solution on network $\calN$ {using only private randomness} as follows. {At each node $v$, from private randomness $Q_v$ generate a random variable $J_v$ uniformly distributed in $[n^2]$, independent of all messages and received signals.} By definition, for each node $u$ for which there exists $\calR\in\scR_{\text{AVC}}(\calN)$ with $R^{(v\to u)}>0$, there is a positive rate solution with arbitrarily small probability of error that conveys data from node $v$ to node $u$. Using these positive rate solutions,  {$J_v$} may be transmitted to all such nodes $u$ essentially for free, because $\log (n^2)$ bits is sub-linear in $n$. Subsequently, all nodes proceed with the original code as if {${\tilQ}_v=q_{vJ_v}$}. Since in the shared randomness coding model, ${\tilQ}_v$ is only available at these nodes $u$, {$J_v$} has been successfully delivered to all the nodes that require it. For state sequence $s^n$, the resulting probability of error is given by
\beq
 \frac{1}{n^{2m}} \sum_{j_1,\ldots,j_m\in[n^2]} e(q_{1j_1},\ldots, q_{mj_m},s^n)
\eeq
which is at most $\eps$ by \eqref{eq:total_derandom}. This proves that the probability of error for {code using only private randomness} can be made arbitrarily small.
\end{IEEEproof}

{
Note that the above argument works equally well for stacked networks; therefore we also have $\tilde\scR_{\AVC}(\underline\calN)=\scR_{\AVC}(\underline\calN)$.
}

\section{Positive Rate {Conditions}}\label{sec:positive}

For both CC and AVC models, it will be important to know whether any information at all can be sent between nodes. This positive (but arbitrarily small) rate will be used for feedback in the CC model and generating shared randomness in the AVC model (see Fig.~\ref{fig:AVC_network}). Thus in this section we investigate the set of node pairs $(u,v)$ for which positive rate can be sent from $u$ to $v$. We do this first without state, and then extend it for the CC and AVC models.

\subsection{Positive Rate Without State}

Assume for now that $\calS$ contains only a single element, in which case $\scR_{\CC}(\calN)=\scR_{\AVC}(\calN)$, and we denote both by $\scR(\calN)$. We form a set $\calP\subset\calV\times\calV$ and subsequently show that $\calP$ is precisely the set of node pairs that can sustain positive rate. For the CC model, we will be interested in whether $(2,1)\in\calP$; \ie whether feedback is possible with respect to the point-to-point channel from node~1 to node~2. On the other hand, for the AVC model, we care whether there exists a node $u$ such that $(u,1),(u,2)\in\calP$.

The set $\calP$ is formed via the following steps:
\begin{enumerate}
\item Initialize $\calP$ as $\{(u,u):u\in\calV\}$. 
\item If there is a pair of nodes $(u,v)\notin\calP$, and a set $\calA\subset\calV$ such that $(j,v)\in\calP$ for all $j\in\calA$, and 
\beq\label{eq:unequal}
\max_{p(x^{(u)}),x^{(\{u\}^c)}} I(X^{(u)};Y^{(\calA)}|X^{(\{u\}^c)}=x^{(\{u\}^c)})>0,
\eeq
then add $(u,v)$ to $\calP$.
\item Repeat step 2 until there are no additional such pairs $(u,v)$.
\end{enumerate}
{Note that the condition in step (2) on a pair of nodes $(u,v)$ is monotonic in the sense that if it holds at any point in the procedure, adding other pairs to $\calP$ cannot cause it to cease holding. Thus, no matter the order in which pairs are added to $\calP$, any pair that satisfies the condition at any point will eventually be added. Thus, the above procedure defines $\calP$ uniquely.}

The mutual information in \eqref{eq:unequal} represents the capacity of a point-to-point channel with input $X^{(u)}$ and output $Y^{(\calA)}$, even though $Y^{(\calA)}$ represents all received values by nodes in $\calA$, which are not available at any single receiver. Additionally, we maximize over constants $\xuc$ in case the channel from $X^{(u)}$ to $Y^{(\calA)}$ only has positive capacity for certain transmissions by the other nodes.

\begin{figure}[htb]
\centerline{\includegraphics[scale=0.7]{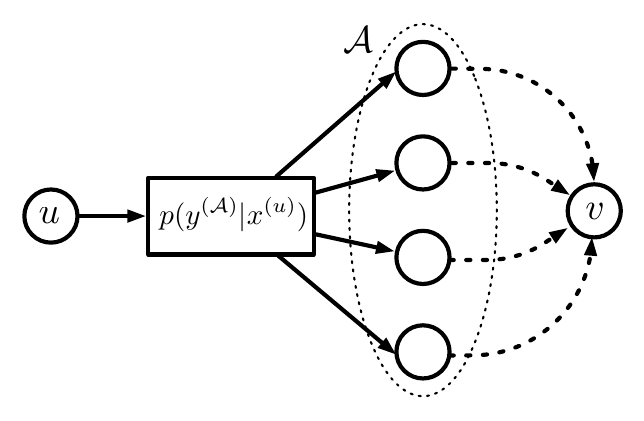}}
\caption{Positive rate can be established between the pair of nodes $(u,v)$
  if  \eqref{eq:unequal} is satisfied and for all $j\in\calA$, positive rate
  can be sent from $j$ to $v$.}
\label{fig:positive_rate}
\end{figure}

\begin{theorem}\label{thm:positive_rate_achievability}
If $(u,v)\in\calP$, then there exists an $\calR\in\scR(\calN)$ with $R^{(u\to v)}>0$.
\end{theorem}
\begin{IEEEproof}
A detailed proof is given by the proof of the stronger result
Lemma~\ref{lemma:state_solutions}, to be stated below. Roughly, the solution
is visualized in Fig.~\ref{fig:positive_rate} and derived as follows. A node may trivially send arbitrary amounts of information to itself; thus $R^{(u\to u)}>0$ is achievable for any $u\in\calV$. We proceed by induction to prove the theorem for pairs $(u,v)\in\calP$ with $u\ne v$. Consider the specific step in the construction of $\calP$ at which $(u,v)$ is added, and let $\calA$ satisfy \eqref{eq:unequal}. We assume that for all $j\in\calA$, positive rate can be sent from $j$ to $v$. To send positive rate from $u$ to $v$, we employ a point-to-point channel code from $X^{(u)}$ to $Y^{(\calA)}$. A message is chosen at node $u$, and the corresponding codeword is transmitted by node $u$ and received by nodes in $\calA$. Next, the received sequences are transmitted from nodes in $\calA$ to node $v$ using positive-rate solutions that are assumed to exist by the induction hypothesis and since by construction $(j,v)\in\calP$ for all $j\in\calA$. Finally, node $v$ decodes the point-to-point code.
\end{IEEEproof}

The following theorem gives the converse result, stating that if $(u,v)\notin\calP$, then values received at node $v$ are conditionally independent of values sent from node $u$ given messages that originate outside node $u$. This indicates that all information known at node $v$ originates outside of node $u$; i.e., the input at node $u$ cannot influence the output at node $v$. This is a much stronger statement than a simple converse, and indeed even stronger than a usual ``strong'' converse, but it is necessary to prove equivalence results.
\begin{theorem}\label{thm:positive_rate_converse}
If $(u,v)\notin\calP$, then for any solution $\calS(\calN)$, $X^{(u)}_{1:n}\to W^{(\{u\}^c\to *)}\to Y^{(v)}_{1:n}$ forms a Markov chain.
\end{theorem}
\begin{IEEEproof}
Fix $(u,v)\notin\calP$. Let $\calA:=\{i:(i,v)\in\calP\}$. By the definition of $\calP$, for any $i\notin\calA$,
\beq
\max_{p(x^{(\{i\})}),x^{(\{i\}^c)}} I(X^{(\{i\})};Y^{(\calA)}|X^{(\{i\}^c)}=x^{(\{i\}^c)})=0.
\eeq
{In other words, the conditional distribution $p(y^{(\calA)}|\bx)$ does not depend on $x^{(\{i\})}$. As this holds for all $i\notin\calA$, it must be that $p(y^{(\calA)}|\bx)=p(y^{(\calA)}|x^{(\calA)})$. Hence,}
for any solution $\calS(\calN)$, we have the Markov chain 
\beq\label{eq:dependency0}
X_t^{(\calA^c)}\to X_t^{(\calA)}\to Y_t^{(\calA)}
\eeq
for each time $t$. We may now write 
\begin{align}
p\left(y^{(\calA)}_{1:n}\Big|w^{(\calA)},x^{(\calA^c)}_{1:n}\right)
&=\prod_{t=1}^n p\left(y^{(\calA)}_t\Big|w^{(\calA)},x^{(\calA^c)}_{1:n},y^{(\calA)}_{1:t-1}\right)
\\&=\prod_{t=1}^n \sum_{x^{(\calV)}_t} p\left(x^{(\calV)}_t\Big|w^{(\calA)},x^{(\calA^c)}_{1:n},y^{(\calA)}_{1:t-1}\right)\,
p\left(y^{(\calA)}_t\Big|x^{(\calV)}_t\right)
\\&=\prod_{t=1}^n \sum_{x^{(\calV)}_t} p\left(x^{(\calV)}_t\Big| w^{(\calA)},x^{(\calA^c)}_{1:n},y^{(\calA)}_{1:t-1}\right)\,
p\left(y^{(\calA)}_t\Big|x^{(\calA)}_t\right)\label{eq:dependency1}
\\&=\prod_{t=1}^n \sum_{x^{(\calA)}_t} p\left(x^{(\calA)}_t\Big| w^{(\calA)},x^{(\calA^c)}_{1:n},y^{(\calA)}_{1:t-1}\right)\,
p\left(y^{(\calA)}_t\Big|x^{(\calA)}_t\right)
\\&=\prod_{t=1}^n \sum_{x^{(\calA)}_t} p\left(x^{(\calA)}_t\Big| w^{(\calA)},y^{(\calA)}_{1:t-1}\right)\,
p\left(y^{(\calA)}_t\Big|x^{(\calA)}_t\right)\label{eq:dependency2}
\\&=\prod_{t=1}^n p\left(y_t^{(\calA)}\Big| w^{(\calA)},y^{(\calA)}_{1:t-1}\right)
=p\left(y^{(\calA)}_{1:n}\Big| w^{(\calA)}\right)
\end{align}
where \eqref{eq:dependency1} follows from \eqref{eq:dependency0}, and \eqref{eq:dependency2} follows by the dependency requirements of the coding at nodes in $\calA$. From this derivation, we conclude that $X^{(\calA^c)}_{1:n}\to W^{(\calA)}\to Y^{(\calA)}_{1:n}$ forms a Markov chain. This completes the proof since $v\in\calA$ and $u\in\calA^c$.
\end{IEEEproof}

 Theorems~\ref{thm:positive_rate_achievability} and~\ref{thm:positive_rate_converse} completely determine when any positive rate is achievable, as stated in the following corollary.
 \begin{corollary}\label{corollary:pos_iff}
 There exists a rate vector $\calR\in\scR(\calN)$ with $R^{(\{v\}\to U)}>0$ if and only if $(v,i)\in\calP$ for all $i\in U$.
 \end{corollary}
 Note that the ``only if'' direction of Corollary~\ref{corollary:pos_iff} is weaker than Theorem~\ref{thm:positive_rate_converse}, because even if $R^{(v\to u)}$ cannot be positive, it does not mean that the strong statement of Theorem~\ref{thm:positive_rate_converse} holds.

\subsection{Positive Rate for the CC Model}

We now extend the above results for CC-type state. For each $s\in\calS$, define $\calP_s$ as above for $\calP$, but with fixed state $S=s$.  Let $\calP_{\CC}=\bigcap_{s\in\calS}\calP_s$.

For any state $s$ such that $(u,v)\in\calP_s$, the following lemma establishes the existence of solutions for the CC model with positive rate from $u$ to $v$ such that (i) if the state is $s$, node~$v$ can reliably decode the message; and (ii) if the state is \emph{not} $s$, node $v$ either decodes correctly or declares an error. Recall that we use the symbol $e$ to signify a decoder declaring an error. We construct these solutions using training sequences (\cf Lemma~\ref{lemma:training}), wherein node $v$ only decodes if $s$ is among the most likely states. Thus if the true state is not $s$, either node $v$ will discover this and declare an error, or the channel is indistinguishable from that with state $s$, so node $v$ will decode reliably. The solutions from this lemma will be used to prove that positive rate can be transmitted from $u$ to $v$ for $(u,v)\in\calP_{\CC}$.

\begin{lemma}\label{lemma:state_solutions}
For any state $s\in\calS$, and all $(u,v)\in\calP_{s}$, there exist a sequence of solutions $\rvS_{u,v,s}^{(n)}(\calN)$ with rate $R^{(u\to v)}>0$ such that
\begin{enumerate}
\item if $S=s$ then the probability of error vanishes with $n$, and 
\item if $S\ne s$ then the probability of making an error without declaring an error  (\ie that $\hatW^{(u\to v)}\notin\{W^{(u\to v)},e\}$) vanishes with $n$.
\end{enumerate}
\end{lemma}
\begin{IEEEproof}
We adopt the convention that a node may send arbitrary amounts of information to itself; thus the lemma is immediate if $u=v$. We proceed by induction to prove the theorem for pairs $(u,v)\in\calP_s$ with $u\ne v$. Consider the specific step in the construction of $\calP_s$ at which $(u,v)$ was added. There is a set $\calA\subset\calV$ such that for some distribution $p(x^{(u)})$ and constant $\xuc$,
\beq\label{eq:mutual_information_state}
I(X^{(u)};Y^{(\calA)}|\Xuc=\xuc,S=s)>0
\eeq
and $(j,v)$ for all $j\in\calA$ has already been added to $\calP_s$. We assume there exist sequences of solutions $\rvS_{j,v,s}^{(n_j)}(\calN)$ for all $j\in\calA$, with rates $R^{(j\to v)}>0$, satisfying the probability of error constraints in the statement of the lemma. Fix a length $n$ to be determined later.

We now describe the coding procedure. Initially node $u$ chooses a message $W^{(u\to v)}\in\calW^{(u\to v)}=[2^{n\tilR^{(u\to v)})}]$, where $\tilR^{(u\to v)}$ is any positive number strictly smaller than the mutual information in \eqref{eq:mutual_information_state}.  Coding proceeds in $3$ sessions, described as follows. The lengths of the first two sessions are $n$, and that of the third session is $\sum_{j\in\calA} n_j$. Thus the quantity $\tilR^{(u\to v)}$ is not the rate achieved by the code, because the overall blocklength is longer than $n$.

\emph{Session 1:} Node $u$ transmits a training sequence $\alpha_{1:n}$ drawn randomly and uniformly from $(\calX^{(u)})^n$ while other nodes transmit the constant $\xuc$. The training sequence constitutes part of the codebook and is revealed to all nodes prior to coding. For each $j\in\calA$, let $Y^{(j)}_{1:n}$ be the received sequence at node $j$ for each $j\in\calA$. 

\emph{Session 2:} Node $u$ transmits $W^{(u\to v)}$ via an $n$-length point-to-point channel code from $X^{(u)}$ to $Y^{(\calA)}$ with input distribution $p(x^{(u)})$ and distribution conditioned on $\Xuc=\xuc$ and $S=s$, while all other nodes transmit the constant $\xuc$. Let $Y^{(j)}_{n+1:2n}$ be the received sequence at node $j$ at each $j\in\calA$.

\emph{Session 3:} Dividing into $|\calA|$ sub-sessions, we run one sub-session for each $j\in\calA$, in which $\rvS_{j,v,s}^{(n_j)}(\calN)$ is employed to transmit $Y^{(j)}_{1:2n}$ from $j$ to $v$, where the blocklength is given by
\beq
n_j=\left\lceil\frac{2n\log|\calY^{(j)}|}{R^{j\to v}}\right\rceil
\eeq
so that $2^{n_jR^{(j\to v)}}\ge |\calY^{(j)}|^{2n}$. Let $\hatY_{1:2n}^{(j)}$ be the decoded sequence at node $v$.

\emph{Decoding:} If any of the solutions $\rvS_{j,v,s}^{(n_j)}(\calN)$ declares an error, then node $v$ declares an error. Otherwise, given $\hatY^{(\calA)}_{1:n}$ node $v$ determines whether $s$ is among the most likely states given the training sequence; that is
\beq\label{eq:argmax} 
{p(\hatY_{1:n}^{(\calA)}|\alpha_{1:n},\Xuc_{1:n}=\xuc_{1:n},S=s)
=\max_{s'}\,  p(\hatY_{1:n}^{(\calA)}|\alpha_{1:n},\Xuc_{1:n}=\xuc_{1:n},S=s').}
\eeq
If \eqref{eq:argmax} does not hold, then node $v$ declares an error. If it does, then node $v$ decodes the message from $\hatY^{(\calA)}_{n+1:2n}$ using the point-to-point channel decoder. Let $\hatW^{(u\to v)}$ be the decoded message.

\emph{Achieved rate:} Recall that all we need to show is that the achieved rate $R^{(u\to v)}$ is positive. The total blocklength for the code is $2n+\sum_{j\in\calA} n_j$, so the overall rate is given by
\begin{align}
R^{(u\to v)}&=\frac{n\tilR^{(u\to v)}}{2n+\sum_{j\in\calA} n_j}\\
&\ge\frac{\tilR^{(u\to v)}}{2+\sum_{j\in\calA} \frac{2\log|\calY^{(j)}|}{R^{j\to v}}+\frac{|\calA|}{n}}.
\end{align}
Note that $R^{(u\to v)}$ is bounded above $0$ for sufficiently large $n$.

\emph{Probability of error analysis:} First consider the case that $S=s$. We need to show that $\Pr(\hatW^{(u\to v)}\ne W^{(u\to v)})$ can be made arbitrarily small. Define the error events
\begin{align}
\calE_1&:=\left\{\hatY^{(\calA)}_{1:2n}\ne Y^{(\calA)}_{1:2n}\right\}\\
\calE_2&:=\left\{s\notin \argmax_{s'}  p(\hatY_{1:n}^{(\calA)}|\alpha_{1:n},\Xuc=\xuc,S=s')\right\}\\
\calE_3&:=\left\{\hatW^{(u\to v)}\ne W^{(u\to v)}\right\}.
\end{align}
The overall error event is $\calE_3$, and we can upper bound its probability by
\beq
\Pr(\calE_3|S=s)\le \Pr(\calE_1|S=s)+\Pr(\calE_1^c\cap\calE_2|S=s)+\Pr(\calE_3|\calE_1^c,\calE_2^c,S=s).
\eeq
By the inductive assumptions that $\rvS_{j,v,s}^{(n_j)}(\calN)$ have vanishing probability of error given state $s$ for each $j\in\calA$, $\Pr(\calE_1)\to 0$. By Lemma~\ref{lemma:training}, $\Pr(\calE_1^c\cap \calE_2|S=s)\to 0$.  Finally, $\Pr(\calE_3|\calE_1^c,\calE_2^c,S=s)$ is merely the probability of error of the point to point code from $u$ to $\calA$, so it vanishes as $n\to\infty$. Thus the overall probability of error may be made arbitrarily small.

Now consider the case that $S=\bars\ne s$. Define the additional error events
\begin{align}
\calE_4&:=\left\{\text{solution }\rvS^{(n_j)}_{j,v,s}(\calN)\text{ declares an error for some } j\in\calA\right\}\\
\calE_5&:=\left\{\hatW^{(u\to v)}\notin\{W^{(u\to v)},e\}\right\}.
\end{align}
We need to show $\Pr(\calE_5)\to 0$ as $n\to\infty$. If either $\calE_2$ or $\calE_4$ occurs, then node $v$ declares an error, so $\calE_5\subset\calE_2^c\cap\calE_4^c$. In addition, $\calE_5\subset\calE_3$, so
\begin{align}
\Pr(\calE_5|S=\bars)
&\le \Pr(\calE_3\cap\calE_2^c\cap\calE_4^c|S=\bars)
\\&\le \Pr(\calE_1\cap\calE_4^c|S=\bars)+\Pr(\calE_3\cap\calE_1^c\cap\calE_2^c\cap\calE_4^c|S=\bars).\label{eq:bars_terms}
\end{align}
The first term in \eqref{eq:bars_terms} vanishes by the inductive assumption on $\rvS_{j,v,s}^{(n_j)}(\calN)$ for all $j\in\calA$. To bound the second term, we consider two cases. First, that $p(y^{(\calA)}|x^{(u)},\xuc,\bars)\ne p(y^{(\calA)}|x^{(u)},\xuc,s)$ for any $x^{(u)}\in\calX^{(u)}$ and $y^{(\calA)}\in\calY^{(\calA)}$. Then $\Pr(\calE_1^c\cap\calE_2^c|S=\bars)\to 0$ by Lemma~\ref{lemma:training}. Otherwise, the channel from $x^{(u)}$ to $Y^{(\calA)}$ conditioned on $\Xuc=\xuc$ is identical for $S=\bars$ and $S=s$. Hence the operation of the point-to-point code from $X^{(u)}$ to $Y^{(\calA)}$ works just as well for $S=\bars$ as for $S=s$, so $\Pr(\calE_3\cap\calE_1^c\cap\calE_2^c|S=\bars)\to 0$.
\end{IEEEproof}

The following theorem gives the positive rate result (equivalent to Theorems~\ref{thm:positive_rate_achievability} and~\ref{thm:positive_rate_converse}) for the CC model.
\begin{theorem}\label{thm:positive_rate_cc}
If $(u,v)\in\calP_{\CC}$, then there exists a rate vector $\calR\in\scR_{\CC}(\calN)$ with $R^{(u\to v)}>0$. Conversely, if $(u,v)\notin\calP_{\CC}$, then for any solution $\calS(\calN)$ there exists $s\in\calS$ such that with $S^n=(s,s,\ldots,s)$, $X^{(u)}_{1:n}\to W^{(\{u\}^c\to *)}\to Y^{(v)}_{1:n}$ forms a Markov chain.
\end{theorem}
\begin{IEEEproof}
To prove the converse, note that if $(u,v)\notin\calP_{\CC}$ then $(u,v)\notin\calP_s$ for some $s\in\calS$. With this fixed state, the proof follows exactly as that of Theorem~\ref{thm:positive_rate_converse}.

Now we prove achievability. Suppose $(u,v)\in\calP_{\CC}$. Thus $(u,v)\in\calP_s$ for all $s\in\calS$. Let $\rvS_{u,v,s}^{(n)}(\calN)$ be the sequence of solutions asserted by Lemma~\ref{lemma:state_solutions}. Let $R^{(u\to v)}_s>0$ be the rate for code $\rvS_{u,v,s}^{(n)}(\calN)$. Let $\tilR^{(u\to v)}=\min_{s\in\calS} R^{(u\to v)}_s$.

We construct a solution to send positive rate from $u$ to $v$ as follows. First node $u$ chooses a message $W^{(u\to v)}\in[2^{n\tilR^{(u\to v)}}]$. Coding proceeds in $|\calS|$ sessions. In the session associated with $s\in\calS$, we employ $\rvS_{u,v,s}^{(n)}(\calN)$ to send $W^{(u\to v)}$ from $u$ to $v$. After all sessions are complete, node $v$ decodes by choosing $\hatW^{(u\to v)}$ to be the output of the first solution that did not declare an error. By Lemma~\ref{lemma:state_solutions}, with high probability the solution associated with the true state will not make an error, and any solution associated with a false state will not make an error without declaring an error. Thus the probability of error is small. As the total blocklength for the code is $n|\calS|$, the achieved rate is $\tilR^{(u\to v)}/|\calS|>0$.
\end{IEEEproof}

\subsection{Positive Rate for the AVC Model}

Recall that, as defined in \cite{CsiszarNarayan:88IT}, an AVC $p(y|x,s)$ is \emph{symmetrizable} if there exists a probability transition matrix $p(s|x)$ such that
\beq\label{eq:symmetrizable}
\sum_{s\in\calS} p(y|x,s) p(s|x') = \sum_{s\in\calS} p(y|x',s) p(s|x),\text{ for all }x,x'\in\calX,y\in\calY.
\eeq
As shown in \cite{CsiszarNarayan:88IT}, a point-to-point AVC has positive capacity if and only if it is non-symmetrizable. Now define $\calP_{\AVC}$ using the same procedure as above for $\calP$, but replace \eqref{eq:unequal} with the condition that there exists $x^{(\{u\}^c)}\in\calX^{(\{u\}^c)}$ such that the channel from $X^{(u)}$ to $Y^{(\calA)}$, conditioned on $X^{(\{u\}^c)}=x^{(\{u\}^c)}$, is non-symmetrizable. 

\begin{theorem}\label{thm:positive_rate_avc}
If $(u,v)\in\calP_{\AVC}$, then there exists a rate vector $\calR\in\scR_{\AVC}(\calN)$ with $R^{(u\to v)}>0$.
\end{theorem}
 \begin{IEEEproof}
 The proof follows from the same argument as for Theorem~\ref{thm:positive_rate_achievability}, except that we replace the point-to-point channel code from $X^{(u)}$ to $Y^{(\calA)}$ with an AVC code. By the assumption that this channel is non-symmetrizable, positive rate can be achieved by the results in \cite{CsiszarNarayan:88IT}.
 \end{IEEEproof}

\section{Compound Channel Equivalence}\label{sec:cc}

In this section and the next we simplify notation by writing $X$ for $X^{(1,1)}$, $Y$ for $Y^{(2,1)}$, and $S$ for $S^{(1)}$. Since we are primarily interested in the independent channel $\calC$, there should be no confusion.

 There are two relevant capacities for the compound channel: first, the standard capacity expression for a compound channel
\beq
\underline{C}=\max_{p(x)}\, \min_{s\in\calS} I(X;Y|S=s),
\label{eq:CC_cap_low}
\eeq
and second, the capacity of a compound channel if the state is known at the encoder and the decoder, wherein the min and max are reversed:
\beq
\bar{C}=\min_{s\in\calS}\, \max_{p(x)} I(X;Y|S=s).
\label{eq:CC_cap_high}
\eeq
{In other words, $\bar{C}$ and $\underline{C}$ represent the capacities of the
independent channel  $\calC$ depending on whether compound state knowledge
is available at the encoder or not.}

Of course, $\underline{C}\le\bar{C}$. Let $\calP_{\CC}$ be defined as above for $\calN$.
As stated in the following theorem, the compound channel is equivalent to a bit-pipe with rate either $\underline{C}$ or $\bar{C}$, depending on whether the rest of the network can sustain any positive feedback rate from node 2 to node 1.
\begin{theorem}
\beq
\scR_{\CC}(\calN)=\begin{cases}
\scR_{\CC}(\calN^{\bar{C}}) & \text{if } (2,1)\in\calP_{\CC}\\
\scR_{\CC}(\calN^{\underline{C}}) & \text{if } (2,1)\notin\calP_{\CC}.\end{cases}
\eeq
\end{theorem}

We prove this theorem in several lemmas, which in combination with continuity from Lemma~\ref{lemma:continuity} prove the theorem. 

\begin{lemma}
{For all networks with links $(2,1)\in \calP_{\CC}$} if $R<\underline{C}$, then $\scR_{\CC}(\calN^R)\subseteq\scR_{\CC}(\calN)$.
\end{lemma}
\begin{IEEEproof}
The proof follows an almost identical argument as that of Lemma~5 from \cite{Equivalence}, which proved that a bit-pipe may simulate a point-to-point noisy channel via a traditional channel code. Recalling that $\underline{C}$ is the usual compound channel capacity, $R<\underline{C}$ implies the existence of a reliable compound channel code at rate $R$. Replacing the channel code in the proof of Lemma~5 from \cite{Equivalence} with such a compound channel code proves our result.
\end{IEEEproof}

\begin{lemma}\label{lemma:cc_achievability2}
{For all networks with links $(2,1)\in \calP_{\CC}$} if $R>\bar{C}$, then $\scR_{\CC}(\calN)\subseteq\scR_{\CC}(\calN^R)$.
\end{lemma}
\begin{IEEEproof}
Let $s^*=\argmin_s \max_{p(x)} I(X;Y)$.
We may use Theorem~6 in \cite{Equivalence}, which proves that a bit-pipe can simulate a noisy channel with less capacity, to simulate the channel $p(y|x,s^*)$ over the bit-pipe of rate $R$, since $R>I(X;Y)$ for this channel and any input distribution.
\end{IEEEproof}

\begin{figure}[t]
\centerline{
\includegraphics[scale=1]{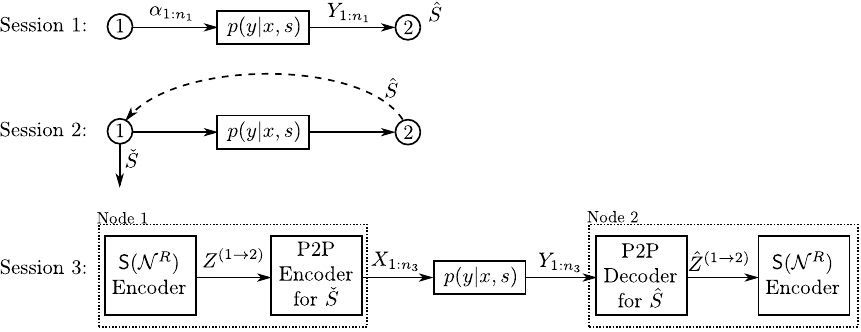}
}
\caption{{The structure of the proof of Lemma~\ref{lemma:CC_training_achievability}. Training is used in Session 1 to learn the state; in Session 2 the estimated state is sent back to the transmitter; in Session 3 a point-to-point channel code is used based on the estimated state.}}
\label{fig:CC_training}
\end{figure}

\begin{lemma}\label{lemma:CC_training_achievability}
{For all networks with links $(2,1)\in \calP_{\CC}$} 
if $R<\bar{C}$, then $\scR_{\CC}(\calN^R)\subseteq\scR_{\CC}(\calN)$.
\end{lemma}
\begin{IEEEproof}
By Theorem~\ref{thm:positive_rate_achievability}, since
$(2,1)\in\calP_{\CC}$, there exists a solution $\mathsf{S}_0(\calN)$ such
that $R^{(2\to 1)}>0$. Given a solution $\mathsf{S}(\calN^R)$, we construct a
solution $\mathsf{S}(\calN)$ with three sessions. In session~1, node~1 sends a training sequence so that node~2 can learn the state. In session~2, this estimated state is transmitted back to node~1 using $\rvS_0(\calN)$. In session 3, node 1 uses this estimated state to transmit a message across $\calC$ while the rest of $\rvS(\calN^R)$ is conducted. {This technique is illustrated in Fig.~\ref{fig:CC_training}.} We give more details as follows.

\emph{Session 1:} We employ a random coding argument wherein we choose a training sequence $\alpha_{1:n_1}$ randomly and uniformly from $\calX^{n_1}$. This sequence forms the codebook for session~1, and it is revealed to nodes 1 and 2. Node~1 transmits $\alpha_{1:n_1}$ into $\calC$ while the inputs to all other channels are arbitrary. Let $Y_{1:n_1}$ be the output of $\calC$. Node~2 forms a state estimate by choosing $\hatS$ arbitrarily from the set {of $\hats\in\calS$ such that $p(Y_{1:n_1}|\alpha_{1:n_1},\hats)=\max_{s'} p(Y_{1:n_1}|\alpha_{1:n_1},s')$}.

\emph{Session 2:} Employ $\rvS_0(\calN)$ with blocklength $n_2$ to transmit $\hatS$ from node~2 to node~1. Let $\check{S}$ be the recovered value at node~1. Assume $n_2$ is large enough such that $2^{n_2R^{(2\to 1)}}\ge |\calS|$.

\emph{Session 3:} {The network conducts $\rvS(\calN^R)$, but signals to be sent along the bit-pipe $\calC^{R}$ are instead transmitted across the noisy link $\calC$ by encoding them at node $1$ using an encoder point-to-point channel with state $\check{S}$, while node~2 employs a decoder for the channel with state $\hatS$.} Let $n_3$ be the blocklength of this session. {Denote by $Z^{(1\to 2)}\in[2^{n_3 R}]$ the signal to be sent across bit-pipe $\calC^R$, and $\hatZ^{(1\to 2)}$ the estimate at node $2$.}

\emph{Probability of error analysis:} Assume the state is $s$. Define the following error events:
\begin{align}
\calE_1&:=\{p(y|x,s)\ne p(y|x,\hatS)\text{ for any }x,y\}\\
\calE_2&:=\{\check{S}\ne\hatS\}\\
\calE_3&:=\{\hatZ^{(1\to 2)}\ne Z^{(1\to 2)}\}.
\end{align}
We may bound the probability of error by
\beq
\Pr(\calE_1)+\Pr(\calE_2)+\Pr(\calE_1^c\cap\calE_2^c\cap\calE_3).
\eeq
By Lemma~\ref{lemma:training}, $\Pr(\calE_1)\to 0$ as $n_1\to\infty$. By Theorem~\ref{thm:positive_rate_achievability}, $\Pr(\calE_2)\to 0$ as $n_2\to\infty$. The effective rate of the point-to-point code in Session~3 is $\frac{nR}{n_3}$, where the total blocklength is $n=n_1+n_2+n_3$. Since by assumption $R<\bar{C}$, for sufficiently large $n_3/(n_1+n_2)$ the effective rate is bounded below $\barC$. Moreover, $\barC\le \max_{p(x)} I(X;Y|S=s)$, so the effective rate is bounded below the capacity of the point-to-point channel with state $s$. As long as $\calE_1$ and $\calE_2$ do not hold, then $\hatS=\check{S}$ are a state for which the operation of the channel is identical to that of $s$, so the channel with this state has the same capacity as with $s$. Hence $\Pr(\calE_1^c\cap\calE_2^c\cap\calE_3)\to 0$ as $n_3\to\infty$.
\end{IEEEproof}

The following theorem is essentially equivalent to Theorem~4 in \cite{Equivalence}, but with a compound channel instead of a standard channel without state.
\begin{lemma}
{For all networks with links $(2,1)\notin \calP_{\CC}$}
if $R>\underline{C}$, then $\scR_{\CC}(\calN)\subseteq\scR_{\CC}(\calN^R)$.
\end{lemma}
\begin{IEEEproof}
By Lemma~\ref{lemma:stacked} it suffices to show that $\scR_{\CC}(\ucalN)\subseteq\scR_{\CC}(\ucalN^R)$. Fix any $\calR\in\text{int}(\scR_{\CC}(\calN))$ and $\lambda>0$. 

\emph{Choose code and define distributions:}
Let $\rvS(\calN)$ be a rate-$\calR$ solution on network $\calN$ for some blocklength $n$. By Theorem~\ref{thm:positive_rate_cc}, for solution $\rvS(\calN)$, $X_{1:n}^{(2)}\to W^{(\{2\}^c\to *)}\to Y_{1:n}^{(1)}$ forms a Markov chain. Moreover, the state $S$ only has direct impact on $Y_{1:n}^{(2)}$, which in turn only has direct impact on $X_{1:n}^{(2)}$. Thus $S\to X_{1:n}^{(2)}\to (W^{(\{2\}^c\to *)},Y_{1:n}^{(1)})$ forms a Markov chain.\footnote{We have written $S$ as a random variable even though it is arbitrary rather than random. By $S\to A\to B$ we mean that $p(b|a,s)=p(b|a)$.} Combining these two chains yields
\beq
S\to X_{1:n}^{(2)}\to W^{(\{2\}^c\to *)}\to Y_{1:n}^{(1)}.
\eeq
Since $W^{(\{2\}^c\to *)}$ is drawn uniformly from $\calW^{(\{2\}^c\to *)}$ and independently from $S$, the distribution of $(W^{(\{2\}^c\to *)},Y_{1:n}^{(1)})$ does not depend on $S$. Thus the distribution of $X_{1:n}^{(1)}$ also does not depend on $S$, as it is a function of $(W^{(\{1\}\to *)},Y_{1:n}^{(1)})$. Therefore, for each time $t$ we may define $p_t(x)$ to be the distribution of $X_t^{(1)}$ independent of $S$. Let $p(x)=\frac{1}{n}\sum_{t=1}^n p_t(x)$ and let
\beq
s^*=\argmin_{s\in\calS} I(X;Y|S=s)
\eeq
where $X$ is drawn from $p(x)$. Let $p_t(x,y)=p_t(x)p(y|x,s^*)$.

\emph{Typical set:} Define $\hatA_{\eps,t}^{(N)}$ to be the $N$-length
typical set according to distribution $p_t(x,y)$ as in \cite[Appendix II]{Equivalence}.

\emph{Design of channel emulators:} By concavity of mutual information with respect to the input variable,
\begin{equation}
\frac{1}{n}\sum_{t=1}^n I(X_t;Y_t|S=s^*)
\le I(X;Y|S=s^*)
=\min_s I(X;Y|S=s)\le \underline{C}<R.
\end{equation}
Let $R_t:=I(X_t;Y_t|S=s^*)+\Delta$ where $\Delta>0$ is chosen so that $\frac{1}{n}\sum_{t=1}^n R_t=R$.

Randomly design decoder $\beta_{N,t}:[2^{NR_t}]\to\ucalY$ by drawing codewords
$\beta_{N,t}(1),\ldots,\beta_{N,t}(2^{NR_t})$ from the i.i.d.~distribution with marginal $p_t(\uy)$.
Define encoder $\alpha_{N,t}:\ucalX\to[2^{NR_t}]$ as
\beq
\alpha_{N,t}(\ux)=\begin{cases}k& \text{if }(\ux,\beta_{N,t}(k))\in\hatA_{\eps,t}^{(N)}\\
1&\text{if }\not\exists k\text{ s.t. }(\ux,\beta_{N,t}(k))\in\hatA_{\eps,t}^{(N)}.\end{cases}
\eeq
Note that the number of bits required to send $(\alpha_{N,t}(\underline{X}))_{t=1}^n$ is $\sum_{t=1}^n NR_t=nNR$, so we may send all these encoded functions via a bit-pipe of rate $R$.

The rest of the proof follows essentially that of Theorem~6 in
\cite{Equivalence}. This involves creating a stacked solution for $\underline\calN$ with exponentially decreasing probability of error, and then converting it into a solution for $\underline\calN^R$ by employing the channel emulators at nodes 1 and 2 to simulate the noisy channel over the rate-$R$ bit-pipe. Finally, the error probability can be bounded provided correct parameters are chosen for the typical set $\hatA_{\eps,t}^{(N)}$, which can be done for our problem by virtue of the fact that $R_t-I(X_t;Y_t|S=s^*)=\Delta>0$.
\end{IEEEproof}

\section{Arbitrarily Varying Channel Equivalence}\label{sec:avc}

The random coding capacity of a point-to-point AVC is defined as the maximum rate that can be achieved if the encoder and decoder have access to shared randomness (inaccessible to the adversary). It is given by
\beq\label{eq:random_coding_capacity}
C_r = \max_{p(x)} \min_{p(s)} I(X;Y).
\eeq
Moreover, the max and min may be interchanged without changing the quantity, because of the convexity properties of the mutual information. Without shared randomness, as shown in \cite{CsiszarNarayan:88IT}, the capacity of an AVC is 0 if the channel is symmetrizable, and $C_r$ if not. Thus, in all cases, $C_r$ is an upper bound on the capacity. The following theorem provides the corresponding network-level converse.

\begin{theorem}\label{thm:avc_converse}
$\scR_{\AVC}(\calN)\subseteq \scR_{\AVC}(\calN^{C_r}).$
\end{theorem}

{The proof of this theorem requires a slightly different approach to network equivalence than that of \cite{Equivalence}. In particular, we use the following Universal Channel Simulation lemma; a version of this result was stated in \cite{Xiang2014} and used for an alternative proof of the network equivalence result. The advantage of this result is that it shows that the difference in distribution (as measured by total variational distance) between a DMC and a simulated channel over a noiseless bit-pipe may be arbitrarily small for any input sequence. That is, no assumptions need to be made on the distribution of the input, which is important because in the AVC setting, this input distribution may be influenced by the adversary, and hence unknown. While \cite{Xiang2014} did not give a complete proof of this lemma, we have provided a proof in Appendix~\ref{appendix_channel_simulation}.\footnote{{
In fact, the result stated in \cite{Xiang2014} is slightly different: it
states that one DMC can be simulated by another; here we only show that a
DMC can be simulated by a noiseless bit-pipe. The result of \cite{Xiang2014}
can be recovered from ours by concatenating an ordinary channel code for the
DMC to be simulated to  the simulation code.}}
\begin{lemma}\label{lemma:channel_simulation}
Consider a DMC $(\calX,q(y|x),\calY)$ with capacity $C$. Given a rate $R>C$,
a noiseless channel simulation code $(f,g)$ consists of
\begin{itemize}
\item $f:\calX^n\times[0,1]\to \{0,1\}^{nR}$,
\item $g:\{0,1\}^{nR}\times[0,1]\to \calY^n$.
\end{itemize}
Let $p(y^n|x^n)$ be the conditional pmf of $Y^n$ given $X^n$ where $Q\sim\text{Unif}[0,1]$ and
\[
Y^n=g(f(X^n,Q),Q).
\]
Let $d_{\text{TV}}(p,q)$ be the total variational distance between two distributions $p$ and $q$. There exists a sequence of length-$n$ channel simulation codes where
\beq
\lim_{n\to\infty} \max_{x^n} d_{\text{TV}}(p(y^n|x^n),q(y^n|x^n))= 0.\label{eq:TV_bound}
\eeq
\end{lemma}
}

\begin{IEEEproof}[Proof of Theorem~\ref{thm:avc_converse}]
{
By the continuity property from Lemma~\ref{lemma:continuity}, it will be enough to show that $\scR_{\AVC}(\calN)\subseteq \scR_{\AVC}(\calN^R)$ for all $R>C_r$. Let
\beq\label{eq:pstar}
p^\star(s):=\argmin_{p(s)} \max_{p(x)} I(X;Y).
\eeq
Let $p^\star(y|x)=\sum_s p^\star(s) p(y|x,s)$. Note that $C_r$ is the capacity of the ordinary channel with transition matrix $p^\star(y|x)$. Since the probability of error for the AVC model is maximized over all choices for $S^n$, it cannot increase if we assume $S^n$ is drawn i.i.d. from $p^\star(s)$. Thus, the capacity region can only enlarge if the AVC is replaced by the ordinary channel $p^\star(y|x)$ in $\calN$. In particular, if we let $\tilde\calN$ be the network in which the AVC is replaced by this channel, we have $\scR_{\AVC}(\calN)\subseteq\scR_{\AVC}(\tilde\calN)$. Thus it will be enough to show $\scR_{\AVC}(\tilde\calN)\subseteq\scR_{\AVC}(\calN^R)$. Moreover, by Lemmas~\ref{lemma:stacked} and~\ref{lemma:avc_randomness} it will be enough to show $\scR_{\AVC}(\underline{\tilde\calN})\subseteq\tilde\scR_{\AVC}(\underline{\calN}^R)$, where as in Sec.~\ref{sec:avc_randomness} $\tilde\scR$ refers to the capacity region under the shared randomness model {from Definition~\ref{def:shared_randomness}}. Take any rate vector $\calR$ in the interior of $\scR_{\AVC}(\underline{\tilde\calN})$, and let $\rvS(\underline{\tilde\calN})$ be a solution with rate vector $\calR$ and probability of error at most $\lambda$. We convert this to a randomized solution on $\underline\calN^R$ as follows. By assumption $R>C_r\ge 0$, so it is certainly possible to transmit data at some positive rate from node 1 to node 2 on network $\underline\calN^R$; thus, by Definition~\ref{def:shared_randomness}, the shared randomness coding model allows arbitrary shared randomness between nodes 1 and 2.
}

{
By Lemma~\ref{lemma:channel_simulation}, for sufficiently large $N$, there exists a length-$N$ channel simulation code $(f,g)$ with rate $R$ where the induced distribution $p(\uy|\ux)$ satisfies
\beq\label{eq:stacked_TV_bound}
\max_{\ux} d_{\text{TV}}\left(p(\uy|\ux),\prod_{\ell=1}^N p^\star(\uy(\ell)|\ux(\ell))\right)\le \lambda/n.
\eeq
Note that in the network $\underline{\tilde\calN}$, $\uX_{1:n}$ and $\uY_{1:n}$ are related by
\beq
p(\uy_{1:n}|\ux_{1:n})=\prod_{t=1}^n \prod_{\ell=1}^N p^\star(\uy_t(\ell)|\ux_t(\ell)).
\eeq
We form a randomized code on network $\underline\calN^R$ by replacing the noisy channel $p^\star(y|x)$ with the channel simulation code used across the layers and repeated $n$ times, once for each time $t\in[n]$. This causes $\uX^n$ and $\uY^n$ to be related by
\beq
\prod_{t=1}^n p(\uy_t|\ux_t).
\eeq
While we have eliminated the state for the channel from node 1 to node 2, the state $s^{(0)}$ for the rest of the network remains. Fix a complete state sequence $\us^{(0)}_{1:n}$, and consider the distribution of random variables
\beq
\uW,\uhatW,\ubX^{(0)}_{1:n},\ubY^{(0)}_{1:n},\uX_{1:n},\uY_{1:n}.
\eeq
conditioned on $\us^{(0)}_{1:n}$. In particular, we wish to bound the total variational distance between the above distribution for the original code on $\underline{\tilde\calN}$, and that for the randomized code on $\underline\calN^R$. Let $\rvP_0$ be the probability law for the distribution of the original code, and for each $t\in[n]$, let $\rvP_t$ be the probability law in which the original noisy channel distribution is replaced by the induced distribution of the channel simulation code for all times $t'\le t$. Thus $\rvP_n$ is the probability law for the code on $\underline\calN^R$, and the difference between $\rvP_{t-1}$ and $\rvP_t$ is only the distribution at time $t$. Using the  generic fact about total variational distance that
\beq
d_{\text{TV}}\big(p(a,b) p(c|b) p(d|a,b,c),\, p(a,b) q(c|b) p(d|a,b,c)\big)
\le \max_{b} d_{\text{TV}}(p(c|b),q(c|b)).
\eeq
we have, for any $t\in[n]$,
\beq
d_{\text{TV}}(\rvP_{t-1},\rvP_t)\le \max_{\ux_t} d_{\text{TV}} \left(p(\uy_t|\ux_t), \prod_{\ell=1}^N p^\star(\uy_t(\ell)|\ux_t(\ell))\right) \le \lambda/n
\eeq
where we have applied \eqref{eq:stacked_TV_bound}. By the triangle inequality,
\beq
d_{\text{TV}}(\rvP_0,\rvP_n)\le \lambda.
\eeq
In particular,
\beq
d_{\text{TV}}(\rvP_0(\uw,\uhatw),\rvP_n(\uw,\uhatw))\le\lambda
\eeq
meaning the probability of error for the randomized code on $\underline\calN^R$ is at most $\lambda$ more than the original probability of error for the code on $\underline{\tilde\calN}$. Note that the state sequence $\us^{(0)}_{1:n}$ affects channel outputs, and thus, via coding operations, may subsequently affect channel inputs. However, because the total variation bound in \eqref{eq:stacked_TV_bound} holds for all input sequences, the effect of the state sequence on the distribution of the channel inputs is irrelevant. Therefore, for any state sequence the resulting randomized code on $\underline{\calN}^R$ has probability of error at most $2\lambda$. Since $\lambda$ may be arbitrarily small, this implies $\calR\in\tilde\scR(\underline{\calN}^R)$.
}
\end{IEEEproof}

Theorem~12.11 from \cite{CsiszarKorner:Book11} states that the capacity of a point-to-point AVC is either 0 or $C_r$. This is shown by proving that a small header can be transmitted from encoder to decoder that allows the encoder and decoder to simulate common randomness. This small header can be sent using any code that achieves positive rate. The following is an extension of this result to the network setting wherein the header may originate at any node and be transmitted to both nodes 1 and 2.

\begin{theorem}\label{thm:avc_equivalence}
If for some node $u$, there exists a rate vector $\calR_1\in\scR_{\AVC}(\calN)$ with $R_1^{(u\to 1)}>0$ and a rate vector $\calR_2\in\scR_{\AVC}(\calN)$ with $R_2^{(u\to 2)}>0$, then $\scR_{\AVC}(\calN)=\scR_{\AVC}(\calN^{C_r})$.
\end{theorem}

\begin{IEEEproof}
In light of Theorem~\ref{thm:avc_converse}, we have only to prove that $\scR(\calN^{C_r})\subseteq \scR(\calN)$. Applying Lemmas~\ref{lemma:stacked}, \ref{lemma:continuity}, and \ref{lemma:avc_randomness}, it is enough to prove $\scR(\underline{\calN}^R)\subseteq \tilde\scR(\underline\calN)$ for all $R<C_r$. By the assumption of the theorem, there exists node $u$ that can transmit data to both nodes 1 and 2; thus by Definition~\ref{def:shared_randomness}, in the shared randomness coding model, $Q_u$ is available at both nodes 1 and 2. By Lemma~12.10 of \cite{CsiszarKorner:Book11}, there exists a randomized point-to-point AVC code achieving any rate $R<C_r$ with arbitrarily small probability of error. Given any solution on $\scR(\underline{\calN}^R)$, we adapt it into a randomized code on $\underline{\calN}$ be employing an $N$-length randomized point-to-point channel code across layers, once for each time $t\in[n]$, using the shared randomness $Q_u$. Since the probability of error of the AVC code is vanishing, for sufficiently large $N$ the probability of the overall code is also vanishing.
\end{IEEEproof}

The following corollary provides a sufficient condition for equivalence for the AVC. It follows immediately from
Theorem~\ref{thm:positive_rate_avc} and Theorem~\ref{thm:avc_equivalence}.
\begin{corollary}\label{avc_result}
If there exists a node $u$ such that $(u,1)\in\calP_{\AVC}$ and $(u,2)\in\calP_{\AVC}$, then $\scR_{\AVC}(\calN)=\scR_{\AVC}(\calN^{C_r})$.
\end{corollary}

\section{AVC Example Network}\label{sec:example}

\begin{figure}
\centerline{\includegraphics[width=.9\textwidth]{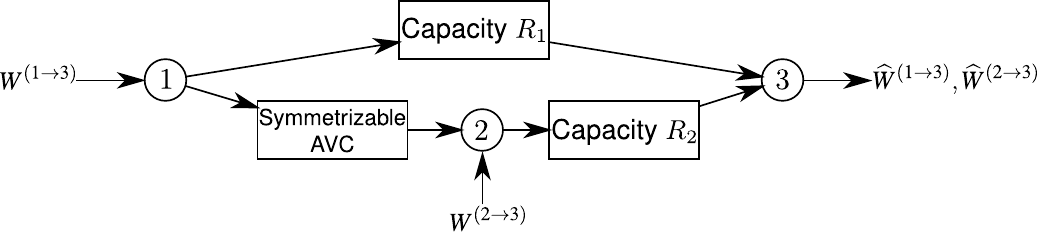}}
\caption{Example network with a symmetrizable AVC from node 1 to node 2 that does not satisfy the conditions of Corollary~\ref{avc_result}. The network also contains a rate $R_1$ bit-pipe between  nodes 1 and 3 and a rate $R_2$ bit-pipe between nodes 2 and 3. Proposition~\ref{prop_example} gives the complete capacity region for this network, which cannot be equated to the capacity region of a network in which the AVC is replaced by any bit-pipe of fixed capacity.}
\label{fig:symmetrizable}
\end{figure}

This section examines the example network shown in Fig.~\ref{fig:symmetrizable}. This network illustrates that when a point-to-point AVC does not satisfy the condition of Corollary~\ref{avc_result}, it is not necessarily equivalent to a zero-capacity bit-pipe, or indeed any bit-pipe with fixed capacity. The channel from node 1 to node 2 is a symmetrizable AVC given by $p(y|x,s)$, with random code capacity $C_r$. The channel from node 1 to node 3 is a bit-pipe with capacity $R_1$, where we assume $R_1>0$, and that from node 2 to node 3 is a bit-pipe with capacity $R_2$. We first determine the capacity region of this network, and then find the capacity region if the AVC were replaced by a bit-pipe of capacity fixed capacity $\tilR$; these two regions do not coincide for any $\tilR$. Roughly, equivalence cannot hold because the symmetrizable AVC leads to a situation in which node 2 can determine that the data sent by node 1 is one of a small number of possibilities. All of these possibilities can be sent along link $(2,3)$, where node 3 can determine which is the correct one using side information from link $(1,3)$. Thus, as long as $R_2$ is not too large, each bit sent on link $(2,3)$ for message $W^{(1\to 3)}$ contributes only a fraction of a bit of useful data; no such phenomenon can occur with a fixed-capacity bit-pipe, since an additional bit would add either a full bit or zero bits to the overall capacity.

It was shown in \cite{HughesAVC97} that with list decoding---even for quite
short lists---the capacity of a symmetrizable AVC is given by its random code capacity. In particular, \cite{HughesAVC97} defines the \emph{symmetrizability} of an AVC $p(y|x,s)$ as the largest integer $M$ for which there exists a stochastic matrix $p(s|x_1,\ldots,x_M)$ such that 
\beq\label{eq:multi_message}
\sum_{s\in \calS} p(y|x,s) p(s|x_1,\ldots,x_M)
\eeq
is symmetric in $x,x_1,\ldots,x_M$. A channel is symmetrizable, in the sense formulated in \cite{CsiszarNarayan:88IT} and discussed above in \eqref{eq:symmetrizable}, if and only if $M\ge 1$. It is shown in \cite{HughesAVC97} that for an AVC with symmetrizability $M$, the decoder can reliably list-decode at rate $C_r$ with list size $M+1$. This result will be instrumental in our examination of the example network.

For the network shown in Fig.~\ref{fig:symmetrizable}, the only positive achievable rates for this network are $R^{(1\to 3)}$ and $R^{(2\to 3)}$. The following proposition characterizes the capacity region for this network.

\begin{proposition}\label{prop_example}
The capacity region for the network shown in Fig.~\ref{fig:symmetrizable} is given by the pairs $(R^{(1\to 3)},R^{(2\to 3)})$ satisfying
\begin{align}
R^{(2\to 3)}&\le R_2\label{eq:region1}\\
R^{(1\to 3)}&\le R_1+C_r\label{eq:region2}\\
R^{(2\to 3)}+(M+1)R^{(1\to 3)}&\le (M+1)R_1+R_2.\label{eq:region3}
\end{align}
\end{proposition}

\begin{IEEEproof}
\emph{Achievability}: The basic idea of our achievability proof is as
follows: node 2 makes use of the list decoding scheme from
\cite{HughesAVC97}, and then transmits along link $(2,3)$ the entire list of
$M+1$ potential messages, in addition to message $W^{2\to 3}$. Along link $(1,3)$, we send part of message $W^{1\to 3}$, in addition to a small hash that
allows node 3 to determine which of the $M+1$ messages is the true one. That
this is possible with a hash of negligible rate is not quite proved by
\cite{HughesAVC97},
since in neither scenario is there a list decoding followed by a
determination of the true message via side information. Here we use a random
linear hash to achieve essentially the same effect as the random choice of
 channel codes in \cite[Lemma 12.8]{CsiszarKorner:Book11}, but in the context of a list code, as we will
 show in the following.

Fix rates $R^{(1\to 3)},R^{(2\to 3)}$ satisfying \eqref{eq:region1}--\eqref{eq:region3}, but with strict inequalities. Fix an integer $q$ and a blocklength $n$. Let $\bbF_{2^q}$ be the finite field of order $2^q$. We express $W^{(1\to 3)}$ as a vector of elements of $\bbF_{2^q}$ as follows. Let $\tilR^{(1\to 3)}$ be the largest multiple of $\frac{q}{n}$ no larger than $R^{(1\to 3)}$. Clearly $\tilR^{(1\to 3)}\ge R^{(1\to 3)}-\frac{q}{n}$. Define integers
\begin{align}
K_1&=\left\lfloor \frac{nR_1}{q}\right\rfloor-1,\label{eq:K1}\\
K_2&=\frac{n\tilR^{(1\to 3)}}{q}-K_1.\label{eq:K2}
\end{align}
By the assumption that $R_1>0$, for $n$ sufficiently large we have $K_1\ge
1$. Message $W^{(1\to 3)}$ is chosen from the alphabet $[2^{n\tilR^{(1\to
    3)}}]$ and message $W^{(2\to 3)}$ from the alphabet $[2^{nR^{(2\to
    3)}}]$, respectively. We may denote $W^{(1\to
  3)}=(W_1,\ldots,W_{K_1+K_2})$ where $W_j\in\bbF_{2^q}$ for all
$j\in[K_1+K_2]$, {where we for the sake of brevity drop the superscript
  $(1\rightarrow 3)$ for  the vector elements}.  Note that the $W_j$ are independent and each drawn uniformly from $\bbF_{2^q}$. For convenience, we write {$W^{K_1+K_2}_{K_1+1}=(W_{K_1+1},\ldots,W_{K_1+K_2})$}.

At the start of encoding, node 1 generates a hash of the vector {$W^{K_1+K_2}_{K_1+1}$}. The symbol $W_1$ is used as the random seed for the hash, and the hash itself is given by
\beq
h=\sum_{j=1}^{K_2} (W_{1})^{j-1} W_{K_1+j}.
\eeq
where $(W_1)^{j-1}$ represents exponentiation in the field $\bbF_{2^q}$. Encoding and decoding proceeds as follows:
\begin{enumerate}
\item $(h,W_{1},\ldots,W_{K_1})$ is transmitted along link $(1,3)$.
\item {$W^{K_1+K_2}_{K_1+1}$} is encoded using an $(M+1)$-list code from \cite{HughesAVC97} and the resulting codeword is transmitted into the AVC $(1,2)$.
\item After receiving the output sequence from the AVC, node 2 decodes the $(M+1)$-length list, denoted ${\hatW^{K_1+K_2}_{i,K_1+1}}=(\hatW_{i,K_1+1},\ldots,\hatW_{i,K_1+K_2})$ for $i\in[M+1]$.
\item $(W^{(2\to 3)},{\hatW^{K_1+K_2}_{i,K_1+1}}:i\in[M+1])$ is transmitted across link $(2,3)$.
\item Node 3 receives the vectors transmitted on links $(1,3)$ and $(2,3)$ without error. It decodes $W^{(2\to 3)}$ from its received vector on link $(2,3)$. Given ${\hatW^{K_1+K_2}_{i,K_1+1}}$ for each $i\in[M+1]$ received on link $(2,3)$, node 3 computes
\beq
\hath_i=\sum_{j=1}^{K_2} (W_{1})^{j-1} \hatW_{i,K_1+j}.
\eeq
For the smallest $i$ for which $\hath_i=h$, node 3 declares
\beq
\hatW^{(2\to 3)}=(W_1,\ldots,W_{K_1},\hatW_{i,K_1+1},\ldots,\hatW_{i,K_1+K_2}).
\eeq
where $h$ and $W_1,\ldots,W_{K_1}$ were received on link $(1,3)$.
\end{enumerate}

\emph{Bit-pipe capacity limits:} We first confirm that in the coding
procedure described above, the vectors sent along links $(1,3)$ and $(2,3)$
do not exceed the capacities of these bit-pipes. The number of bits sent
along link $(1,3)$ is $(K_1+1)q\le nR_1$, so its capacity constraint is
satisfied. 

{From \eqref{eq:K2} we obtain
\beq
K_2q 
=n\tilR^{(1\to 3)}-K_1q
\le n\tilR^{(1\to 3)}-nR_1+2q
\le nR^{(1\to 3)}-nR_1+2q
\label{eq:K_2q}
\eeq where the first inequality is due to $qK_1\le nR_1-2$ from
\eqref{eq:K1}.
Using the r.h.s.~from \eqref{eq:K_2q}, the number of bits sent along link}
$(2,3)$ is now given as  
\beq
(M+1)K_2q+nR^{(2\to 3)}
\le n(M+1)R^{(1\to 3)}-n(M+1)R_1+2q+nR^{(2\to 3)}.
\eeq
Since \eqref{eq:region3} holds with a strict inequality, this quantity is at most $nR_2$ for sufficiently large $n$.

\emph{Probability of error:}  There are two potential sources of error: (i)
the decoded list from the AVC at node 2 does not include the true intended
message, and (ii) there exists $i\in[M+1]$ such that $\hath_i=h$ even though
${\hatW^{K_1+K_2}_{i,K_1+1}\ne W^{K_1+K_2}_{K_1+1}}$. 
For the first source of error, note that the
number of bits in ${W^{K_1+K_2}_{K_1+1}}$
is $K_2q$, so the rate of the list code on the
AVC 
{can be obtained from \eqref{eq:K_2q} as 
\beq
\frac{K_2q}{n} 
=\tilR^{(1\to 3)}-\frac{K_1q}{n}
\le R^{(1\to 3)}-R_1+\frac{2q}{n}\label{eq:AVC_bits}
\eeq}
Since \eqref{eq:region2} holds with a strict inequality,
the quantity on the {l.h.s.}~in \eqref{eq:AVC_bits} is less than $C_r$ for sufficiently
large $n$. Thus, by the results in \cite{HughesAVC97}, the probability that the decoded list does not include the true message vanishes with $n$.

Now consider the second source of error. The content of the decoded list
depends only on ${W_{K_1+1}^{K_1+K_2}}$, the state sequence $S^n$, and
the random operation of the AVC. In particular, the list is independent of
$W_{1}$. Thus, for any 
{$w_{K_1+1}^{K_1+K_2},\hatw_{K_1+1}^{K_1+K_2}\in\bbF_{2^q}^{K_2}$}
\beq
{\Pr(\hath_i=h|W_{K_1+1}^{K_1+K_2}=w_{K_1+1}^{K_1+K_2},\hatW_{K_1+1}^{K_1+K_2}=\hatw_{K_1+1}^{K_1+K_2})}
=\Pr\left(\sum_{j=1}^{K_2} (W_{1})^{j-1} (\hatw_{j}-w_j)=0\right).
\eeq
If {$w_{K_1+1}^{K_1+K_2}\ne \hatw_{K_1+1}^{K_1+K_2}$} 
then the polynomial in $W_1$ inside the probability is a nonzero polynomial
of degree at most $K_2-1$, so it has at most $K_2-1$ roots. Since
$W_{K_2+1}$ is chosen uniformly from $\bbF_{2^q}$, if
{$w_{K_1+1}^{K_1+K_2}\ne \hatw_{K_1+1}^{K_1+K_2}$}
\beq
{\Pr(\hath_i=h|W_{K_1+1}^{K_1+K_2}=w_{K_1+1}^{K_1+K_2},\hatW_{K_1+1}^{K_1+K_2}=\hatw_{K_1+1}^{K_1+K_2})}\le\frac{K_2-1}{2^q}.
\eeq
Therefore, the probability that $\hath_i=h$ for any $i$ satisfying
${\hatW^{K_1+K_2}_{i,K_1+1}\ne W^{K_1+K_2}_{K_1+1}}$ 
is at most
\beq
\frac{(K_2-1) M}{2^q}.
\eeq
This can be made arbitrarily small for sufficiently large $q$.

\emph{Converse}: Let $(R^{(1\to 3)},R^{(2\to 3)})$ be an achievable rate pair. Thus there exists a sequence of solutions $\rvS_n(\calN)$ of length $n$, rates $R^{(1\to 3)},R^{(2\to 3)}$ and probability of error going to $0$ as $n\to\infty$. In this argument, we use the fact that the capacity region does not change if the state $S^n$ is chosen randomly, as long as this random choice is independent of the message (but it may depend on the code). We consider two specific distributions for $S^n$ under $\rvS_n(\calN)$ for some $n$.  First, that $S^n$ is chosen randomly from the i.i.d. distribution with marginal $p^\star(s)$ defined in \eqref{eq:pstar} as the saddle-point in the random coding capacity. With this choice, the AVC behaves as a (stateless) stationary memoryless channel with transition probability
\beq
p^\star(y|x)=\sum_{s} p^\star(s) p(y|x,s).
\eeq
Note that the channel $p^\star(y|x)$ has capacity $C_r$. Simple applications of the cutset bound yield \eqref{eq:region1} and \eqref{eq:region2}.

To prove \eqref{eq:region3}, we consider a different distribution on the state. Let $p_{X^n}(x^n)$ be the distribution of the input sequence to the AVC $(1,2)$ under solution $\rvS_n(\calN)$. Note that this distribution depends only on the code at node 1, so it is independent of the state $S$ of the AVC. The state sequence $S^n$ is drawn from the distribution
\beq
\sum_{x_1^n,\ldots,x_M^n} p_{X^n}(x_1^n)\cdots p_{X^n}(x_M^n) \prod_{i=1}^n p(s_i|x_{1i},\ldots,x_{Mi})
\eeq
where the distribution $p(s|x_1,\ldots,x_m)$ is one for which
\eqref{eq:multi_message} is symmetric. {{Let $z_1\in [2^{nR_1}]$ and
    $z_2\in [2^{nR_2}]$ with the corresponding random variables $Z_1$ and $Z_2$} denote the input symbols of links $(1,3)$ and $(2,3)$ respectively. Since these links are bit-pipes, these variables also represent the output symbols of the respective links. We also write $X^n$ and $Y^n$ for the input and output sequences of the AVC $(1,2)$. We may now write the distribution of all relevant random variables, conditioned on state sequence $S^n=s^n$, by
\begin{multline}
p(w^{(1\to 3)},w^{(2\to 3)}, x^n,y^n,z_1,z_2,\hatw^{(1\to 3)},\hatw^{(2\to 3)}|s^n)
\\
\begin{aligned}=\,&\frac{1}{2^{nR^{(1\to 3)}} 2^{nR^{(2\to 3)}}}
\, {p(x^n|w^{(1\to 3)})
\, p(z_1|w^{(1\to 3)})}
\\&\cdot \left[\prod_{i=1}^n p(y_i|x_i,s_i)\right]
\,
{p(z_2|y^n,w^{(2\to 3)})}
\\ &\cdot
{p(\hatw^{(1\to 3)} | z_1,z_2)
\,
p(\hatw^{(2\to 3)} | z_1,z_2)}
\end{aligned}
\end{multline}
{where the encoding and decoding operations are written as conditional distributions because randomized coding is allowed.}
}
{
Let $V(y|x,x_1,\ldots,x_M)$ be the symmetric distribution in \eqref{eq:multi_message}. The distribution of $X^n,Y^n$ may be written as
\beq
p_{X^n}(x^n)
\sum_{x_1^n,\ldots,x_M^n} p_{X^n}(x_1^n)\cdots p_{X^n}(x_M^n) \prod_{i=1}^n V(y_i|x_i,x_{1i},\ldots,x_{Mi}).
\eeq
Thus, the distribution of $X^n,Y^n$ is unchanged if we let $X_1^n,\ldots,X_M^n$ be random sequences, each distributed according to $p_{X^n}$, and independent from each other, from $X^n$, and from the messages, and where $Y^n$ is drawn from
\beq
\prod_{i=1}^n V(y_i|x_{1i},\ldots,x_{Mi}).
\eeq
This induces a probability law on all variables other than $S^n$ given by
\begin{multline}
p(w^{(1\to 3)},w^{(2\to 3)}, x^n,x_1^n,\ldots,x_M^n,y^n,z_1,z_2,\hatw^{(1\to 3)},\hatw^{(2\to 3)})
\\
\begin{aligned}
=\,& \frac{1}{2^{nR^{(1\to 3)}} 2^{nR^{(2\to 3)}}}
\, {p(x^n|w^{(1\to 3)})
\, p(z_1|w^{(1\to 3)})}
\\&\cdot p_{X^n}(x_1^n)\cdots p_{X^n}(x_M^n) \left[\prod_{i=1}^n V(y_i|x_{1i},\ldots,x_{Mi})\right]
\\&\cdot {p(z_2|y^n,w^{(2\to 3)})}
{p(\hatw^{(1\to 3)} | z_1,z_2)
\, p(\hatw^{(2\to 3)} | z_1,z_2)}.
\end{aligned}
\end{multline}
Note in particular that}
$X^n,X_1^n,\ldots,X_M^n,Y^n$ are distributed according to
\beq
p_{X^n}(x^n)p_{X^n}(x_1^n)\cdots p_{X^n}(x_M^n) \prod_{i=1}^n V(y_i|x_i,x_{1i},\ldots,x_{Mi}).
\eeq
By Fano's inequality and the data processing inequality,
\beq\label{eq:R23}
nR^{(2\to 3)}\le I(W^{(2\to 3)};Z_2)+n\eps_n
\eeq
where $\eps_n\to 0$ as $n\to\infty$. Applying Fano's inequality again, we have
\begin{align}
nR^{(1\to 3)}&=H(W^{(1\to 3)})
\\&\le I(W^{(1\to 3)};Z_1,Z_2)+n\eps_n
\\&= I(W^{(1\to 3)};Z_2)+I(W^{(1\to 3)};Z_1|Z_2)+n\eps_n
\\&\le I(W^{(1\to 3)};Z_2)+nR_1+n\eps_n
\\&\le I(X^n;Z_2)+nR_1+n\eps_n\label{eq:R13a}
\end{align}
where in \eqref{eq:R13a} we have used the fact that $W^{(1\to 3)}\to X^n\to Z_2$ is a Markov chain. By symmetry of 
{$(X^n,X_1^n,\ldots,X_M^n)$, we have $I(X_k^n;Z_2)=I(X^n;Z_2)$ for all $k\in[M]$. Thus, defining $\eps'_n=(M+2)\eps_n$ and $X_0^n=X^n$,}
{
\begin{align}
&nR^{(2\to 3)}+(M+1) n R^{(1\to 3)}
\\&\le  I(W^{(2\to 3)};Z_2)+\sum_{k=0}^M I(X_k^n;Z_2)+(M+1)nR_1+n\eps'_n
\\&\le I(W^{(2\to 3)};Z_2)+\sum_{k=0}^M I(X_k^n;Z_2|W^{(2\to 3)},X_0^n,\ldots,X_{k-1}^n)+(M+1)nR_1+n\eps'_n\label{eq:mutual_independence}
\\&= I(W^{(2\to 3)},X_0^n,\ldots,X_M^n;Z_2)+(M+1)nR_1+n\eps'_n
\\&\le n R_2 + (M+1)nR_1+n\eps'_n
\end{align}}%
where in \eqref{eq:mutual_independence} we have used the fact that $(W^{(2\to 3)},X_0^n,\ldots,X_M^n)$ are mutually independent. Dividing by $n$ and taking the limit as $n\to\infty$ yields \eqref{eq:region3}.
\end{IEEEproof}

Suppose that in the example network the AVC were replaced by a bit-pipe of capacity $\tilR$. It is easy to see that the resulting set of achievable $(R^{(1\to 3)},R^{(2\to 3)})$ pairs is given by
\begin{align}
R^{(2\to 3)}&\le R_2\\
R^{(1\to 3)}&\le R_1+\tilR\\
R^{(1\to 3)}+R^{(2\to 3)}&\le R_1+R_2.
\end{align}
This region does not correspond to \eqref{eq:region1}--\eqref{eq:region2} for any value of $\tilR$, as long as $M\ge 1$ (i.e., the AVC is symmetrizable). Therefore, the AVC in Fig.~\ref{fig:symmetrizable} is not equivalent to any fixed capacity bit-pipe.

\section{Relation to the ``Edge Removal'' Problem}\label{sec:edge_removal}

Consider two networks $\mathcal{N}$ and $\mathcal{N}'$ with identical
topologies except for a single edge, which has capacity $C_e$ in network
$\mathcal{N}$, but capacity $C_e'=C_e-\delta$ in network
$\mathcal{N}'$. Herein, $\delta>0$ is a small constant. Particular attention
has been devoted recently to the so called \textit{edge removal} problem which describes
the special case of this scenario for $C_e=\delta$.  It has been shown in
\cite{HEJ10,JEH11} that for a variety of demand types for which the network
coding capacity can be described by the cut-set bound, the capacity of every
cut is reduced by at most $\delta$ for each dimension. This means that if
a rate vector $\mathcal{R}$ is achievable in network $\mathcal{N}$, a rate
vector  $\mathcal{R}-\delta \mathcal{I}$ is achievable in $\mathcal{N}'$,
where $\mathcal{I}$ denotes the unit rate vector. Examples include
single and multisource multicast and single source cases with non-overlapping
demands, but also scenarios for which the cut-set bound is not tight, for
example a specific class of  multiple unicast networks
\cite{JEH11}. Further, in \cite{LE11} the edge removal
problem has also been connected to the problem whether a network coding
instance allows a reconstruction with  $\epsilon$ and zero error,
respectively. However, so far only various special cases have been
considered, and it is not clear how to formulate the edge removal problem for
general demands and topologies. 

\begin{figure}[htb]
\centerline{\includegraphics[scale=0.6]{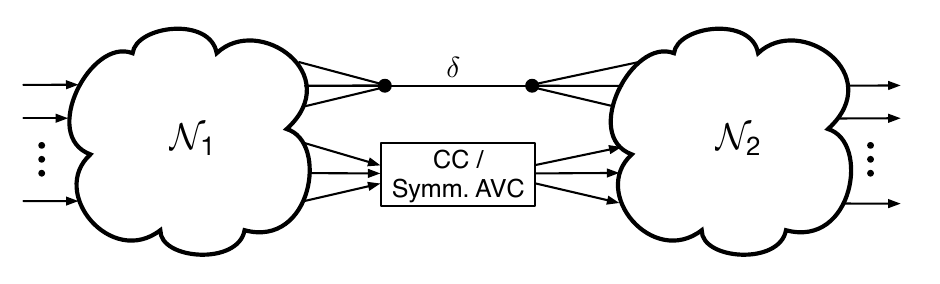}}
\caption{The network $\mathcal{N}$ consists of two arbitrary networks
  $\mathcal{N}_1$ and  $\mathcal{N}_2$ connected by an edge with capacity
  $\delta>0$ and a CC or alternatively, a symmetrizable AVC.}
\label{fig:EdgeRemoval}
\end{figure}

In the following, based on the discussion in
Sections~\ref{sec:cc} and~\ref{sec:avc}, we extend the edge removal problem to networks with
state. We formulate our result for both the CC and the AVC case in the following theorem.
\begin{theorem}
Given a network $\mathcal{N}$ with state according to \eqref{eq:model} and
assume that a non-zero rate vector $\mathcal{R}(\mathcal{N})$
is achievable. Further, assume
that there exists a single edge  with capacity $\delta$ in the
network. {Let the network  $\mathcal{N}'$ be defined as the network
$\mathcal{N}$ with the $\delta$-capacitated edge removed.}
Then, there exists a network {$\mathcal{N}$ such that for the corresponding edge-removed network $\mathcal{N}'$,}
{$\mathcal{R}(\mathcal{N'})<\mathcal{R}(\mathcal{N})-\delta\, \mathbf{I}$,
where $\mathbf{I}$ denotes the identity matrix. }
\end{theorem}
\begin{IEEEproof}
  We show this by considering the example in Fig.~\ref{fig:EdgeRemoval},
  where two networks $\mathcal{N}_1$ and $\mathcal{N}_2$ are connected via a
  CC or a symmetrizable point-to-point AVC, resp., and an edge of
  capacity $\delta$. Suppose that this connection also represents the
  min-cut of the overall network $\mathcal{N}$. For the AVC case, as the
  capacity of the symmetrizable AVC is either $0$ or $C_r$, removing the
  $\delta$-capacitated edge leads to a network capacity of
  $\mathcal{R}_{\text{AVC}}(\mathcal{N'})=0$ according to
  Theorem~\ref{thm:avc_equivalence}. For the CC case the capacity of the CC
  is either $\underline{C}$ or $\bar{C}$ (see \eqref{eq:CC_cap_low} and
  \eqref{eq:CC_cap_high}). By removing the $\delta$-capacitated feedback
  edge the network capacity is reduced from
  $\mathcal{R}_{\text{CC}}(\mathcal{N})=\bar{C}$ to
  $\mathcal{R}_{\text{CC}}(\mathcal{N'})=\underline{C}$, where
  $\bar{C}-\underline{C}$ can be larger than $\delta$.
\end{IEEEproof}

\section{Conclusion}\label{sec:conclusion}
We have considered reliable communication over noisy network in the presence of active adversaries. This is
modeled by a subset of independent point-to-point channels consisting of AVCs or CCs. For these cases
we have identified scenarios for which the capacity of the corresponding
noisy state-dependent network equals the capacity of another state-less
network in which the AVCs or CCs are replaced by noiseless bit-pipes. Our
results indicate that, in the network setting, the equivalent capacity of
these channels is not necessarily equal to their capacity in an isolated
point-to-point scenario. For example, the point-to-point AVC represents a
 pessimistic model for the action of an adversary, leading to zero
capacity in some cases. We have shown that in a network setting such a
pessimistic model becomes much more optimistic and leads to a positive rate
if additional network connectivity exits between the head and the tail node
of the AVC or CC under consideration. As most modern communication is performed in an
underlying networking framework, this suggests that existing results may be insufficient for characterizing networks in the presence of active adversaries.

\appendices

\section{Proof of Lemma~\ref{lemma:channel_simulation}}\label{appendix_channel_simulation}

We make use of the method of types, adopting notation from \cite{CoverThomas:Book}. Specifically, given a sequence $x^n$, define its type as
\beq
P_{x^n}(x) = \frac{|\{i:x_i=x\}|}{n}.
\eeq
Similarly define the joint type of a pair of sequences $(x^n,y^n)$ as $P_{x^n,y^n}$. Given a type $P_X$, define the type class $T(P_X)$ as the set of sequences $x^n$ with $P_{x^n}=P_X$.

Fix $\eps>0$, and define $\tilde\calP_n$ to be the set of $n$-length types $P_{XY}$ such that
\beq
|P_{XY}(x,y)-P_{X}(x) q(y|x)|\le \eps P_X(x) q(y|x)\text{ for all }x\in\calX,y\in\calY
\eeq
{where $q(y|x)$ is the channel to be simulated}. Observe that $P_{x^n,y^n}\in\tilde\calP_n$ if and only if $(x^n,y^n)$ is robustly typical  \cite{OrlitskyRoche2001} with respect to the distribution $P_{x^n}(x) q(y|x)$. By the Conditional Typicality Lemma from Chapter 2 of \cite{ElGamalKim:11Book}, since $x^n$ is trivially robustly typical with respect to $P_{x^n}$ (indeed, with parameter $\eps=0$), if $Y^n\sim \prod_{i=1}^n q(y_i|x_i)$, then with probability approaching 1, $P_{x^n,Y^n}\in\tilde\calP_n$.

Let $I(P_X,P_{Y|X})$ be the mutual information between $X$ and $Y$ where $(X,Y)\sim P_X P_{Y|X}$. By continuity of mutual information, for any $\gamma>0$, there exists $\eps$ small enough so that for all $P_{XY}\in \tilde\calP_n$,
\beq
I(P_X,P_{Y|X})\le I(P_X,q(y|x))+\gamma\le C+\gamma.
\eeq
In particular, if we choose $\gamma=(R-C)/2$, then for sufficiently small $\eps$,
\beq
I(P_X,P_{Y|X})\le R-\gamma.
\eeq

We construct a noiseless channel simulation code out of a number of codebooks, one for each type $P_{XY}\in\tilde\calP_n$. A codebook of joint type $P_{XY}$, denoted $\calC(P_{XY})$, is a subset of $T(P_Y)$. We say a codebook with joint type $P_{XY}$ is \emph{feasible} if, for all $x^n\in T(P_X)$, there exists a sequence $y^n\in \calC(P_{XY})$ where $P_{x^n,y^n}=P_{XY}$. Define
\beq\label{eq:M_P_def}
M= 2^{n(R-\delta)}
\eeq
where $0<\delta<\gamma$. We claim that for sufficiently large $n$, for all $P_{XY}\in\tilde\calP_n$ there exists a feasible codebook of size at most $M$. To prove this, consider a random choice of codebook $\calC(P_{XY})$ consisting of $M$ sequences chosen uniformly and independently from $T(P_Y)$. {Note that the codebook will contain fewer than $M$ unique sequences if the same sequence is chosen more than once.} We show that with positive probability this codebook is feasible. For each $x^n\in T(P_X)$ define the event
\beq
\calE(x^n):=\{P_{x^n,y^n}\ne P_{XY}\text{ for all }y^n\in\calC(P_{XY})\}.
\eeq
Note that the only random variable in this event is the codebook itself. Define the conditional type class
\beq
T_{P_{XY}}(x^n):=\{y^n:P_{x^n,y^n}=P_{XY}\}.
\eeq
Note that
\beq
\calE(x^n)=\{T_{P_{XY}}(x^n)\cap \calC(P_{XY})=\emptyset\}.
\eeq 
By using standard bounds on the size of type classes, for any $x^n\in T(P_X)$
\beq
\frac{|T_{P_{XY}}(x^n)|}{|T(P_Y)|}\ge\frac{1}{(n+1)^{|\calY|-1}} 2^{-nI(P_X,P_{Y|X})}.
\eeq
For any $x^n\in T(P_X)$, we may bound the probability of event $\calE(x^n)$ by
\begin{align}
\bbP(\calE(x^n))
&= \left(1-\frac{|T_{P_{XY}}(x^n)|}{|T(P_Y)|}\right)^{M}
\\&\le \left(1-\frac{1}{(n+1)^{|\calY|-1}} 2^{-nI(P_X,P_{Y|X})}\right)^{M}
\\&\le \exp\left\{-\frac{1}{(n+1)^{|\calY|-1}} M 2^{-nI(P_X,P_{Y|X})}\right\}.
\end{align}
Thus, by the union bound
\begin{align}
\bbP\left(\bigcup_{x^n\in T(P_X)}\calE(x^n)\right)
&\le |\calX|^n \exp\left\{-\frac{1}{(n+1)^{|\calY|-1}} M 2^{-nI(P_X,P_{Y|X})}\right\}
\\&= |\calX|^n \exp\left\{-\frac{1}{(n+1)^{|\calY|-1}} 2^{n(R-I(P_X,P_{Y|X})-\delta)}\right\}
\\&\le |\calX|^n \exp\left\{-\frac{1}{(n+1)^{|\calY|-1}} 2^{n(\gamma-\delta)}\right\}
\end{align}
This quantity is vanishing in $n$ since $\delta<\gamma$, so for sufficiently large $n$ there exists at least one feasible codebook $\calC(P_{XY})$ of size at most $M$.

We now describe a channel simulation code. Assume $n$ is large enough such there exists at least one feasible codebook for each $P_{XY}\in\tilde\calP_n$. 

\emph{Encoder:} Given input sequence $x^n$, randomly choose a sequence 
\beq
\tilY^n\sim \prod_{i=1}^n q(y_i|x_i).
\eeq
Let $P_{XY}=P_{x^n,\tilY^n}$. If $P_{XY}\in\tilde\calP_n$, randomly choose a codebook $\calC(P_{XY})$ uniformly from among all feasible codebooks of size at most $M$ for this type. If $P_{XY}\notin\tilde\calP_n$, declare an error. Of the sequences $y^n\in\calC(P_{XY})\cap T_{P_{XY}}(x^n)$ (there must be at least one, since the codebook is feasible), choose one uniformly at random, which we denote $Y^n$. The encoder outputs two bit-strings:
\begin{enumerate}
\item A string of length $\lceil \log |\tilde\calP_n|\rceil$ denoting the type $P_{XY}$.
\item A string of length $\log M$ denoting the index of $Y^n$ in $\calC(P_{XY})$.
\end{enumerate}
Note that {for sufficiently large $n$,} the total number of bits is at most $nR$, since $|\tilde\calP_n|\le 2^{n\gamma}$ for sufficiently large $n$ and $\gamma>0$. 

\emph{Decoder:} Upon learning $P_{XY}$, the decoder can determine the chosen feasible codebook $\calC(P_{XY})$, since it has access to the same randomness as the encoder, and thus it can recover $Y^n$.

To bound the variational distance, we first note that, for a given joint type $P_{XY}$, if there is at least one feasible codebook of size at most $M$, then each sequence $y^n\in T(P_Y)$ appears in exactly the same number of such codebooks. Indeed, consider two sequences $y_1^n,y_2^n\in T(P_Y)$. There exists a permutation that takes $y_1^n$ to $y_2^n$. Applying this permutation to the codebook preserves feasibility, because both the input type class $T(P_X)$ and the output type class $T(P_Y)$ are unchanged by permutation. Thus, the permutation constitutes a bijection between feasible codebooks containing $y_1^n$ and feasible codebooks containing $y_2^n$. This implies that they are equal in number. Thus, if $P_{XY}=P_{x^n,\tilY^n}\in\tilde{\calP}_n$, then the randomly chosen codebook $\calC(P_{XY})$ is equally likely to contain any sequence $y^n\in T_{P_{XY}}(x^n)$, and hence $Y^n$ is uniformly distributed among $T_{P_{XY}}(x^n)$. Hence, for any pair of sequences $x^n,y^n$ where $P_{x^n,y^n}\in\tilde{\calP}_n$, the induced distribution from the simulation code is given by
\beq\label{eq:induced_distribution}
p(y^n|x^n) = \bbP(P_{x^n,\tilY^n}=P_{x^n,y^n}) \frac{1}{|T_{P_{x^n,y^n}}(x^n)|}.
\eeq

Now, for the discrete memoryless channel $q(y|x)$, the probability $q(y^n|x^n)$ depends only on the joint type of $(x^n,y^n)$. Thus, conditioning on a particular joint type, the output sequence is uniformly distributed among the conditional type class. In other words, the right-hand side of \eqref{eq:induced_distribution} is precisely equal to $q(y^n|x^n)$ for all $x^n,y^n$. Since \eqref{eq:induced_distribution} only holds if  $P_{x^n,y^n}\in\tilde{\calP}_n$, the total variational distance between $p(y^n|x^n)$ and $q(y^n|x^n)$ is at most the probability that $P_{x^n,\tilY^n}\notin\tilde{\calP}_n$, which, as argued above, vanishes as $n\to\infty$.


\end{document}